\documentclass[twocolumn]{aastex631}

\usepackage{amsmath}
\usepackage{multirow}
\usepackage{soul}
\usepackage{bm}

\newcommand{\Loss}{\mathcal{L}}
\newcommand{\NN}{\mathcal{N}}
\newcommand{\xx}{\mathbf{x}}
\newcommand{\xr}{\mathbf{\hat{x}}}
\newcommand{\xp}{\mathbf{x'}}
\newcommand{\pp}{\mathbf{p}}
\newcommand{\zz}{\mathbf{z}}
\newcommand{\uu}{\mathbf{u}}
\newcommand{\fx}{\mathbf{f}}
\newcommand{\gz}{\mathbf{g}}
\newcommand{\bz}{\mathbf{b}}
\newcommand{\sig}{\bm{\sigma}}
\newcommand{\mask}{\bm{\mathcal{M}}}

\shorttitle{A Probabilistic Autoencoder for Type Ia Supernovae}

\shortauthors{Stein et al.}
\graphicspath{{./}{figures/}}

\begin{document}

\title{A Probabilistic Autoencoder for Type Ia Supernovae Spectral Time Series}

\author[0000-0002-5193-516X]{George Stein}
\correspondingauthor{George Stein}
\email{gstein@berkeley.edu}
\affiliation{Berkeley Center For Cosmological Physics, University of California, Berkeley, Campbell Hall 341 Berkeley, CA 94720}
\affiliation{Physics Division, Lawrence Berkeley National Laboratory, 1 Cyclotron Road, Berkeley, CA, 94720}

\author{Uro\v{s} Seljak}
\affiliation{Berkeley Center For Cosmological Physics, University of California, Berkeley, Campbell Hall 341 Berkeley, CA 94720}
\affiliation{Physics Division, Lawrence Berkeley National Laboratory, 1 Cyclotron Road, Berkeley, CA, 94720}

\author[0000-0003-3801-1912]{Vanessa B\"ohm}
\affiliation{Berkeley Center For Cosmological Physics, University of California, Berkeley, Campbell Hall 341 Berkeley, CA 94720}
\affiliation{Physics Division, Lawrence Berkeley National Laboratory, 1 Cyclotron Road, Berkeley, CA, 94720}

\author{G.~Aldering}
\affiliation{Physics Division, Lawrence Berkeley National Laboratory, 1 Cyclotron Road, Berkeley, CA, 94720}

\author[0000-0002-0389-5706]{P.~Antilogus}
\affiliation{Laboratoire de Physique Nucl\'eaire et des Hautes Energies, CNRS/IN2P3, \\ Sorbonne Universit\'e, Universit\'e de Paris, 4 place Jussieu, 75005 Paris, France}

\author[0000-0002-9502-0965]{C.~Aragon}
\affiliation{Physics Division, Lawrence Berkeley National Laboratory, 1 Cyclotron Road, Berkeley, CA, 94720}
\affiliation{College of Engineering, University of Washington 371 Loew Hall, Seattle, WA, 98195}

\author{S.~Bailey}
\affiliation{Physics Division, Lawrence Berkeley National Laboratory, 1 Cyclotron Road, Berkeley, CA, 94720}

\author[0000-0003-0424-8719]{C.~Baltay}
\affiliation{Department of Physics, Yale University, New Haven, CT, 06250-8121}

\author{S.~Bongard}
\affiliation{Laboratoire de Physique Nucl\'eaire et des Hautes Energies, CNRS/IN2P3, \\ Sorbonne Universit\'e, Universit\'e de Paris, 4 place Jussieu, 75005 Paris, France}

\author[0000-0002-5828-6211]{K.~Boone}
\affiliation{Physics Division, Lawrence Berkeley National Laboratory, 1 Cyclotron Road, Berkeley, CA, 94720}
\affiliation{Department of Physics, University of California Berkeley, 366 LeConte Hall MC 7300, Berkeley, CA, 94720-7300}
\affiliation{DIRAC Institute, Department of Astronomy, University of Washington, 3910 15th Ave NE, Seattle, WA 98195, USA}

\author[0000-0002-3780-7516]{C.~Buton}
\affiliation{Univ Lyon, Universit\'e Claude Bernard Lyon~1, CNRS/IN2P3, IP2I Lyon, F-69622, Villeurbanne, France}
   
\author[0000-0002-5317-7518]{Y.~Copin}
\affiliation{Univ Lyon, Universit\'e Claude Bernard Lyon~1, CNRS/IN2P3, IP2I Lyon, F-69622, Villeurbanne, France}

\author[0000-0003-1861-0870]{S.~Dixon}
\affiliation{Physics Division, Lawrence Berkeley National Laboratory, 1 Cyclotron Road, Berkeley, CA, 94720}
\affiliation{Department of Physics, University of California Berkeley, 366 LeConte Hall MC 7300, Berkeley, CA, 94720-7300}

\author[0000-0002-7496-3796]{D.~Fouchez}
\affiliation{Aix Marseille Univ, CNRS/IN2P3, CPPM, Marseille, France}

\author[0000-0001-6728-1423]{E.~Gangler}  
\affiliation{Univ Lyon, Universit\'e Claude Bernard Lyon~1, CNRS/IN2P3, IP2I Lyon, F-69622, Villeurbanne, France}
\affiliation{Universit\'e Clermont Auvergne, CNRS/IN2P3, Laboratoire de Physique de Clermont, F-63000 Clermont-Ferrand, France}

\author[0000-0003-1820-4696]{R.~Gupta}
\affiliation{Physics Division, Lawrence Berkeley National Laboratory, 1 Cyclotron Road, Berkeley, CA, 94720}

\author[0000-0001-9200-8699]{B.~Hayden}
\affiliation{Physics Division, Lawrence Berkeley National Laboratory, 1 Cyclotron Road, Berkeley, CA, 94720}
\affiliation{Space Telescope Science Institute, 3700 San Martin Drive Baltimore, MD, 21218}

\author{W.~Hillebrandt}
\affiliation{Max-Planck-Institut f\"ur Astrophysik,  Karl-Schwarzschild-Str. 1, D-85748 Garching, Germany}

\author[0000-0003-2495-8670]{Mitchell Karmen}
\affiliation{Physics Division, Lawrence Berkeley National Laboratory, 1 Cyclotron Road, Berkeley, CA, 94720}

\author[0000-0001-6315-8743]{A.~G.~Kim}
\affiliation{Physics Division, Lawrence Berkeley National Laboratory, 1 Cyclotron Road, Berkeley, CA, 94720}

\author[0000-0001-8594-8666]{M.~Kowalski}
\affiliation{Institut f\"ur Physik,  Humboldt-Universitat zu Berlin, Newtonstr. 15, 12489 Berlin}
\affiliation {DESY, D-15735 Zeuthen, Germany}

\author[0000-0002-9207-4749]{D.~K\"usters}
\affiliation {Department of Physics, University of California Berkeley, 366 LeConte Hall MC 7300, Berkeley, CA, 94720-7300}
\affiliation {DESY, D-15735 Zeuthen, Germany}

\author[0000-0002-8357-3984]{P.-F.~L\'eget}
\affiliation{Laboratoire de Physique Nucl\'eaire et des Hautes Energies, CNRS/IN2P3, \\ Sorbonne Universit\'e, Universit\'e de Paris, 4 place Jussieu, 75005 Paris, France}

\author{F.~Mondon}  
\affiliation{Universit\'e Clermont Auvergne, CNRS/IN2P3, Laboratoire de Physique de Clermont, F-63000 Clermont-Ferrand, France}

\author[0000-0001-8342-6274]{J.~Nordin}
\affiliation{Physics Division, Lawrence Berkeley National Laboratory, 1 Cyclotron Road, Berkeley, CA, 94720}
\affiliation{Institut f\"ur Physik,  Humboldt-Universitat zu Berlin, Newtonstr. 15, 12489 Berlin}

\author[0000-0003-4016-6067]{R.~Pain}
\affiliation{Laboratoire de Physique Nucl\'eaire et des Hautes Energies, CNRS/IN2P3, \\ Sorbonne Universit\'e, Universit\'e de Paris, 4 place Jussieu, 75005 Paris, France}

\author{E.~Pecontal}
\affiliation{Centre de Recherche Astronomique de Lyon, Universit\'e Lyon 1, 9 Avenue Charles Andr\'e, 69561 Saint Genis Laval Cedex, France}

\author{R.~Pereira}
\affiliation{Univ Lyon, Universit\'e Claude Bernard Lyon~1, CNRS/IN2P3, IP2I Lyon, F-69622, Villeurbanne, France}

\author[0000-0002-4436-4661]{S.~Perlmutter}
\affiliation{Physics Division, Lawrence Berkeley National Laboratory, 1 Cyclotron Road, Berkeley, CA, 94720}
\affiliation{Department of Physics, University of California Berkeley, 366 LeConte Hall MC 7300, Berkeley, CA, 94720-7300}

\author[0000-0002-8207-3304]{K.~A.~Ponder}
\affiliation{Department of Physics, University of California Berkeley, 366 LeConte Hall MC 7300, Berkeley, CA, 94720-7300}

\author{D.~Rabinowitz}
\affiliation{Department of Physics, Yale University, New Haven, CT, 06250-8121}

 \author[0000-0002-8121-2560]{M.~Rigault}
\affiliation{Univ Lyon, Universit\'e Claude Bernard Lyon~1, CNRS/IN2P3, IP2I Lyon, F-69622, Villeurbanne, France}

\author[0000-0001-5402-4647]{D.~Rubin}
\affiliation{Department of Physics and Astronomy, University of Hawai`i, 2505 Correa Rd, Honolulu, HI, 96822}
\affiliation{Physics Division, Lawrence Berkeley National Laboratory, 1 Cyclotron Road, Berkeley, CA, 94720}

\author{K.~Runge}
\affiliation{Physics Division, Lawrence Berkeley National Laboratory, 1 Cyclotron Road, Berkeley, CA, 94720}

\author[0000-0002-4094-2102]{C.~Saunders}
\affiliation{Physics Division, Lawrence Berkeley National Laboratory, 1 Cyclotron Road, Berkeley, CA, 94720}
\affiliation{Department of Physics, University of California Berkeley, 366 LeConte Hall MC 7300, Berkeley, CA, 94720-7300}
\affiliation{Princeton University, Department of Astrophysics, 4 Ivy Lane, Princeton, NJ, 08544}
\affiliation{Sorbonne Universit\'es, Institut Lagrange de Paris (ILP), 98 bis Boulevard Arago, 75014 Paris, France}

\author[0000-0002-9093-8849]{G.~Smadja}
\affiliation{Univ Lyon, Universit\'e Claude Bernard Lyon~1, CNRS/IN2P3, IP2I Lyon, F-69622, Villeurbanne, France}

\author{N.~Suzuki}   
\affiliation{Physics Division, Lawrence Berkeley National Laboratory, 1 Cyclotron Road, Berkeley, CA, 94720}
\affiliation{Kavli Institute for the Physics and Mathematics of the Universe, The University of Tokyo Institutes for Advanced Study, \\ The University of Tokyo, 5-1-5 Kashiwanoha, Kashiwa, Chiba 277-8583, Japan}

\author{C.~Tao}
\affiliation{Tsinghua Center for Astrophysics, Tsinghua University, Beijing 100084, China}
\affiliation{Aix Marseille Univ, CNRS/IN2P3, CPPM, Marseille, France}

\author[0000-0002-4265-1958]{S.~Taubenberger}
\affiliation{Max-Planck-Institut f\"ur Astrophysik, Karl-Schwarzschild-Str. 1, D-85748 Garching, Germany}

\author{R.~C.~Thomas}
\affiliation{Physics Division, Lawrence Berkeley National Laboratory, 1 Cyclotron Road, Berkeley, CA, 94720}
\affiliation{Computational Cosmology Center, Computational Research Division, \\ Lawrence Berkeley National Laboratory, 1 Cyclotron Road, Berkeley, CA, 94720}

\author{M.~Vincenzi}
\affiliation{Physics Division, Lawrence Berkeley National Laboratory, 1 Cyclotron Road, Berkeley, CA, 94720}
\affiliation{Institute of Cosmology and Gravitation, University of Portsmouth, Portsmouth, PO1 3FX, UK}

\collaboration{50}{The Nearby Supernova Factory}





\begin{abstract}
We construct a physically-parameterized probabilistic autoencoder (PAE) to learn the intrinsic diversity of type Ia supernovae (SNe~Ia) from a sparse set of spectral time series. The PAE is a two-stage generative model, composed of an Auto-Encoder (AE) which is interpreted probabilistically after training using a Normalizing Flow (NF). We demonstrate that the PAE learns a low-dimensional latent space that captures the nonlinear range of features that exists within the population, and can accurately model the spectral evolution of SNe~Ia across the full range of wavelength and observation times directly from the data. By introducing a correlation penalty term and multi-stage training setup alongside our physically-parameterized network we show that intrinsic and extrinsic modes of variability can be separated during training, removing the need for the additional models to perform magnitude standardization. We then use our PAE in a number of downstream tasks on SNe~Ia for increasingly precise cosmological analyses, including automatic detection of SN outliers, the generation of samples consistent with the data distribution, and solving the inverse problem in the presence of noisy and incomplete data to constrain cosmological distance measurements. We find that the optimal number of intrinsic model parameters appears to be three, in line with previous studies, and show that we can standardize our test sample of SNe~Ia with an RMS of $0.091 \pm 0.010$ mag, which corresponds to $0.074 \pm 0.010$ mag if peculiar velocity contributions are removed. Trained models and codes are released at \href{https://github.com/georgestein/suPAErnova}{github.com/georgestein/suPAErnova}.
\end{abstract}



\section{Introduction}
\label{sec:intro}

Type Ia supernovae (SNe~Ia) are excellent probes of cosmic history, leading to the discovery of the accelerating expansion of the universe \citep{sn1, sn2}. Their cosmological utility emerges from the high degree of similarity between each SN~Ia, and differences in their luminosity have been shown to strongly correlate with observed spectro-temporal features. As such, they can be used as ``standardizable candles'' to infer the relative distances to them through a measurement of their fluxes, which alongside a measurement of their redshift allows for the expansion history of the universe to be inferred \citep{sn1, sn2, sn3, sn4}.

The limiting factor in using SNe~Ia for increasingly precise cosmological analyses is a detailed understanding of their spectral diversity and evolution, which cannot be modelled from first principles to high enough accuracy. Thus the field relies on data-driven models, which have uncovered a number of well-known relations between SNe~Ia features and luminosity, including the correlation between the peak luminosity of a SN~Ia and the light curve decrease time \citep[brighter-slower effect;][]{phillips1993}, and the dependence of the peak luminosity on color \citep[the brighter-bluer effect][]{riess1996, tripp1998}. These behaviors are captured in conventional light curve fitting routines such as SALT2~\citep{salt2}, MLCS2k2 \citep{mlcs2007}, and SNooPy \citep{snoopy2011}, and their associated standardization parameters.

While these effects correlate with a high degree of spectral variation, they are insufficient to fully account for the detailed differences between spectral and temporal features of different SNe~Ia. To try to understand spectral behavior in more detail, collections of SN~Ia spectra \citep[e.g.,][]{matheson2008, bailey2009, silverman2012, folatelli2013, stahl2020} have been used to examine the strengths, ratios, and velocities of specific spectral features \citep[e.g.,][]{nugent1995, folatelli2004, branch2006, arsenijevic2008, bailey2009, folatelli2010, foley2011, blondin2012, silverman2012b, folatelli2013, wang2013} and correlate these with SNe~Ia brightness, color, and decline rate.

With the advent of full SN~Ia spectral time series \citep{aldering2020} or their virtually-constructed analogs \citep{siebert2019, stahl2020}, it has become possible to study the full spectro-temporal behavior of SNe~Ia. Examples include the first construction of a spectral metric space \citep{sasdelli2015}, Gaussian Process twinning \citep{fakhouri2015}, expectation maximization factor analysis \citep{snemo}, spectral feature factor analysis \citep{sugar}, a SN~Ia autoencoder \citep{SN_autoencoder} hierarchical Bayesian spectro-temporal modeling \citep{bayesianSED}, deep learning \citep{stahl2020b}, and the non-linear Twins Embedding space of \citet{twins1, twins2}.

Such models must be able to account for both extrinsic and intrinsic modes of spectral diversity. Intrinsic effects result from object-to-object differences between supernovae explosions, while extrinsic effects are differences caused by physical processes external to the supernovae system. Examples of extrinsic effects include the amount of Galactic and extra-galactic dust along the sightline to the object (and hence extinction), and the peculiar velocity of the supernovae with respect to our observational rest frame, which should therefore be uncorrelated with the intrinsic properties of the SN~Ia. Depending on the specific empirical model, they can be used for applications including magnitude standardization, anomaly detection, and uncertainty estimates.

\subsection{Empirical modelling of SNe~Ia}
 After the initial explosion a supernovae continues to brighten until it reaches a peak and begins to fade, with the observable life-cycle (for the purpose of this work) taking on the order of $\sim$50 days. Accurately modelling this supernovae luminosity as a function of time is challenging due to both the small number of spectroscopically observed SNe~Ia and the highly irregular time sampling of spectra from each object, where for most supernovae we observe only $\sim$10 spectra spread over the range. This sparse time sampling can prove difficult for numerical techniques that require more uniform observations, so fitting an empirical model of supernova flux as a function of time and wavelength often has two steps. The first is to interpolate the observations from each supernova onto a more regularly spaced time grid, generally achieved through spline interpolations or Gaussian Processes. The second is to then model the spectrum at each temporal location and wavelength bin on this time grid. The most commonly used models are based on variations of a principal component analysis (PCA), and schematically take a form that separates intrinsic and extrinsic physical effects into unique terms: 
\begin{equation}
\begin{split}
    \mathrm{Flux}_{SN}(p, \lambda) &= \mathrm{Amplitude}_{\mathrm{SN}} \\ &\times \left [F_0(p, \lambda) + x_{1,\mathrm{SN}} * F_1(p, \lambda) + ... \right]\\&\times c_{\mathrm{SN}}*\mathrm{Extinction}(\lambda),
\end{split}
\end{equation}
where $p$ is the time from peak brightness and $\lambda$ is the rest-frame wavelength. The average spectral sequence is described by $F_0(t, \lambda)$, the components that describe additional PCA variability are $F_n(t, \lambda)$, where $n>0$, and the color term representing both extinction and intrinsic color variations are properties of the global model. The parameters indicated with a $\mathrm{SN}$ subscript are fit to to each supernova and correspond to PCA amplitudes $x_{1,\mathrm{SN}}$ and the extinction $c_{\mathrm{SN}}$. The terms of the model are split this way for a few key reasons:

{\textit{Amplitude:}} For each supernova the observed redshift $z_{obs}$ is generally well known, such that the wavelength and time scales can be accurately de-redshifted. The observed redshift has contributions from the peculiar velocity of the supernovae which are extrinsic to the supernova explosion. A leading amplitude term then ensures that the peculiar velocity component is not correlated with the model parameters, and that it is the only coefficient dependent on the flux normalization. This amplitude can interchangeably be written as $10^{-0.4 \Delta M}$ when working in magnitudes.

{\textit{Color law:}} The color law attempts to account for dust along the line-of-sight to the supernovae. The optical depth to each supernova involves a number of factors, including corrections from the local environment of the supernovae and its host galaxy, and line-of-sight variations along the inter-galactic medium and within our own galaxy. The Milky Way extinction can be determined independently and removed from the observed spectrum. Any optical depth variations along the line of sight should be slowly varying with respect to the $\sim$50~day observation window of a supernovae \citep{huang2017}, and therefore should be dependent only on the wavelength of observation. This color law is generally an input to the model \citep{salt2, snemo, bayesianSED, twins1}, and any time-dependent color variation should be captured by the variations to the global model. 

To use such an empirical model for magnitude standardization then requires an additional third step, in which the (possible) correlations between model parameters and intrinsic luminosity are uncovered. This requires an additional model to be fit to ``explain'' the magnitude residual as a function of the model parameters from previous steps. 

In this work we propose an alternative to this three step workflow -- a probabilistic autoencoder (PAE) to model supernovae spectra as a function of observation time. As introduced by \citet{PAE}, a PAE combines the advantages of an Auto-Encoder (i.e. it is fast and easy to train) with the desired properties of a generative model, which makes a PAE a powerful tool for probabilistic data reconstruction and outlier detection of SNe~Ia. We physically parameterize our PAE and introduce a multi-stage training setup and correlation penalty term in order to separate and decorrelate intrinsic and extrinsic effects during training, which removes the need for an additional model to perform magnitude standardization. 

Our method has a number of advantages over PCA-based models. First, the autoencoder has the ability to learn complicated non-linear mappings between the best fit latent representations (parameters), while a PCA analysis is limited to linear transformations. This allows for increased spectral diversity to be expressed over linear models for a given latent dimensionality. Second, the probabilistic nature of the PAE allows for a straight-forward determination of outlying spectra and calculation of the errors on the best-fit model parameters within the observational errors.
A conditional autoencoder (AE) can account for time-evolution by simply feeding in the observation times as a conditional parameter, and does not need to first pre-process the data to interpolate it onto a regular grid, which allows the model to work directly on the data. Finally, a PAE
model can be used to generate artificial SNe~Ia samples consistent with the data distribution, and to create a faithful simulation of SNe~Ia spectro-temporal series. 

The outline of this paper is as follows. We first describe the dataset and reference baselines in Section~\ref{sec:data}, followed by a detailed description of probabilistic autoencoders in Section~\ref{sec:PAE}. We then outline our architecture and training setup in Section~\ref{sec:architecture}. Section~\ref{sec:results} showcases the PAE results, where we demonstrate that the PAE provides better fits to the observations than the most commonly used model in the literature, it automatically detects outliers, and provides an accurate fit on supernovae parameters and their errors. A discussion follows in Section~\ref{sec:discussion}.

\section{Dataset \& Reference Baselines}
\label{sec:data}
Our dataset consists of spectral time series data of 228 unique SNe~Ia, obtained by the Nearby Supernova Factory \citep[SNfactory;][]{aldering2002, aldering2020} using the SuperNova Integral Field Spectrograph \citep[SNIFS;][]{lantz2004}. The original spectra span the range 3200--10000\,\AA\ simultaneously. The spectra from SNIFS were reduced using the SNfactory data reduction pipeline \citep{bacon2001, aldering2006, scalzo2010}, flux calibrated following \citet{buton2013, rubin2022}, and host-galaxy subtracted as in \citet{bongard2011}. The spectra were corrected for dust in our Galaxy using the dust map from \citet{schlegel98} and the
extinction-color relation from \citet{cardelli89}.

Following our past procedure for similar analyses \citep{fakhouri2015, snemo, sugar, aldering2020, twins1, twins2}, the wavelength and phases have been transformed to the restframe, and the fluxes have been transformed to a reference redshift of $z = 0.05$ using the appropriate factors of $z$ and $1+z$ using redshifts from \citet{childress2013} and \citet{Rigault2020}. Because SN Ia spectral features are broad, the spectra are rebinned to a common restframe wavelength binning of 1000~km/s between 3300 -- 8600~\AA,  resulting in $N_\lambda=288$ restframe wavelength bins. Each spectrum is accompanied by an uncertainty spectrum, $\sigma_{SN}(\lambda)$. A small number of spectra do not cover all wavelength bins, therefore for each spectrum we construct a mask array $\mathcal{M}_{SN}(\lambda)$ to flag any missing wavelength bins.

Each supernova has between 5 to 64 observations at different times, for a total of 3034 spectra. The time gaps between each observation are typically in the range of 2-3 days at early phases and longer at later phases, but with exceptions due to, e.g., bad weather. The given observation time $p$ is the phase relative to the peak luminosity of the supernovae in B-band as fit by the SALT2 model~\citep{salt2}, in days. The SALT2 fits also report an uncertainty on the time on peak luminosity. We cut data outside of ($-10$ days, $+40$ days), resulting in 2696 final spectra for a minimum of 4 observations of a supernovae, to a maximum of 32. The amplitudes of the spectra, initially in the $z=0.05$ reference frame, are multiplied by a constant to scale the range of values to $\sim$(0,1).

 
 At a given observation time, spectra from different supernovae have a high degree of similarity, and it is easy to imagine each unique spectra being described by a set of modifications to some mean spectral envelope as a function of time. It has been shown before that the leading few components in a PCA analysis capture a significant amount of the supernovae-to-supernovae variation \citep{salt2, snemo}, so we expect the data to be able to be represented by an autoencoder with some small set of latent variables. 
 
\subsection{Reference Baselines}
\label{sec:baselines}
Throughout this work we compare our spectral reconstructions to the SALT2 model \citep{salt2}, and compare our cosmological distance measurements to both the SALT2 model and the Twins Embedding \citep{twins1, twins2}.

\subsubsection{SALT2}
\label{sec:salt2}
SALT2 \citep{salt2} models the time-evolving spectral energy distribution as 
\begin{align}
\begin{split}
    \label{eq:salt}
    F_{SN}(p, \lambda) &= x_{0, \mathrm{SN}} \left[ M_0(p,\lambda)+x_{1, \mathrm{SN}} M_1(p, \lambda) + ... \right] \\ 
    & \hspace{0.5cm} \times \mathrm{exp}[c_{\mathrm{SN}}\, \mathrm{CL}(\lambda)],
\end{split}
\end{align}
where $p$ is the rest-frame time since the date of maximum luminosity in B-band and $\lambda$ is the rest-frame wavelength. The $M_0$ component is the average spectral sequence, $M_i$ for $i>0$ are additional components that describe further object-to-object variability, and $\mathrm{CL(\lambda)}$ is a generic color term that mixes dust extinction and intrinsic color variations left over after decorrelating $x_1$ and $c$. Each individual supernovae is then parameterized by a combination of these components multiplied by leading amplitude terms describing the strength of each: $x_{i, SN}$, and $c_{SN}$. $x_{0,SN}$ is the flux normalization and is a function of both the intrinsic luminosity and the luminosity distance of the supernovae. The best-fit SALT2 parameters $x_{0, SN}$, $x_{1, SN}$, and $c_{SN}$ were fit for each light curve in the dataset, and we used {\texttt {sncosmo}} \citep{sncosmo} to generate the best fit restframe SALT2 spectra for each supernova at each observation time. 

The SALT2 light curve fits are used to determine the peak brightness of each SN~Ia, and then a linear correction for the light curve width and color is applied to ``explain'' the magnitude residual to each object as a function of the other model parameters:
\begin{equation}
    \label{eq:salt_standardization}
    M_{res, \mathrm{SN}}= M_{B,\mathrm{SN}} - M_{ref} + \alpha x_{1, \mathrm{SN}} - \beta c_{\mathrm{SN}}.
\end{equation}

The arbitrary reference magnitude $M_{ref}$ and standardization parameters $\alpha$ and $\beta$ are fit in order to minimize the magnitude residual $M_{res, i} \equiv \Delta M^{SALT2}$.

\subsubsection{Twins Embedding}
\label{sec:twins}
The Twins Embedding \citep{twins1, twins2} does not model temporal evolution, and instead aims to explain the spectral variability of SNe~Ia at maximum light. There are four separate components to the model:
\begin{enumerate}
    \item A differential time evolution model to estimate a spectrum at maximum light for each SN~Ia.
    \item A second ``Reading Between the Lines'' (RBTL) model to fit for a mean spectrum at maximum light, $f_{mean}(\lambda)$, and explain the supernovae-to-supernovae variability at maximum light as a function of two parameters $\Delta M_i$ and $\Delta A_{V,i}$,
    \begin{equation}
    f_{\mathrm{model},i}(\lambda) = f_{\mathrm{mean}}(\lambda) \times 10^{-0.4(\Delta M_i+\Delta A_{V,i} CL(\lambda))}.
    \end{equation}
    $\Delta M_i$ is the difference in intrinsic brightness compared to the mean spectrum in magnitudes, and $\Delta A_{V,i}$ represents the coefficient of the extinction-color relation that best matches the supernova’s spectrum to the mean function. The RBTL model is used to deredden each spectrum at maximum light to remove extrinsic contributions from distance uncertainties and interstellar dust.
    \item A third non-linear ``Twins Embedding'' model is trained on the dereddenned spectra from the RBTL model in order to further explain any variability of SN~Ia spectra at maximum light. The Twins Embedding uses the Isomap algorithm \citep{Isomap} to embed the spectral distance \begin{equation}
        \gamma_{i,j} = \sqrt{ \sum_k \left( \frac{f_{dered, i}(\lambda_k) - f_{dered, j}(\lambda_k)}{f_{mean}(\lambda_k)} \right)^2}
    \end{equation}
    between two SNe~Ia labeled i and j into a low-dimensional (3D) space $\xi$ while preserving the distances between nearby points in the high-dimensional space. 
    \item Gaussian Process (GP) regression is then used to infer the magnitude residuals ($\Delta M_i$ from step 2) of SNe~Ia over the 3D Twins Embedding space $\bm{\xi}$, and to reconstruct spectra from a given embedding vector. The inferred value of the magnitude residual can be subtracted from the measured value, and the remainder represents the ``unexplained residual''
\end{enumerate}

Dixon et al. (in prep) extends this Twins Embedding to the full time series using a neural network.

\section{Physically Parameterized Probabilistic Autoencoder}
\label{sec:PAE}

\begin{figure*}
\centering
\includegraphics[width=1.0\textwidth]{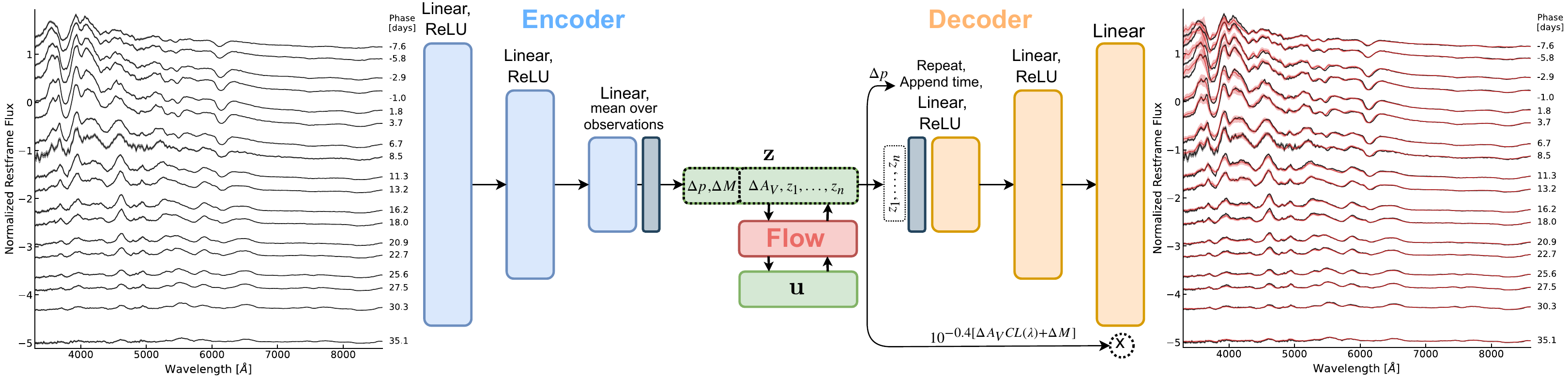}
\caption{Probabilistic autoencoder architecture. The encoder receives as input the observed spectra $\xx$ and corresponding observation times $\pp$, and extracts a set of time-independent latent parameters $\zz$ for each supernova. The decoder combines the latent parameters with the desired observation times to reconstruct the data. Both are fully connected neural networks, consisting of a chain of linear layers each followed by a ReLU activation. By separating out the flow of certain latent parameters through the decoder, along with the addition of a correlation penalty during training, we explicitly inform the model to learn physically motivated latent parameters expressing extrinsic ($\Delta p$, $\Delta A_V$) and intrinsic ($z_1, ... , z_n$) modes of variability. After the encoder-decoder is trained, a normalizing flow learns a bijective mapping between the unconstrained $\zz$ space and a Gaussian latent space $\uu$, which allows for the determination of the supernovae density in comparison to others in the dataset.}
\label{fig:AE}
\end{figure*}

Our probabilistic autoencoder is constructed in two separate stages. First, we train a conditional autoencoder to learn a low-dimensional latent representation of each supernova that is independent of the observation time(s). After the autoencoder is trained we construct a normalizing flow to map from the unconstrained autoencoder latent space to a Gaussian latent space. For clarity, our data notation is as follows, where for each supernova, $SN$, arrays are filled sequentially using the observed spectra:
\begin{itemize}
    \item $\xx_{N_{\mathrm {SN}} \times N_{\mathrm {obs}} \times N_\lambda}$: Observed SNe~Ia spectral time series.
    \item $\xr_{N_{\mathrm {SN}} \times N_{\mathrm {obs}} \times N_\lambda}$: Reconstructed SNe~Ia spectral time series.
    \item $\sig_{N_{\mathrm {SN}} \times N_{\mathrm {obs}} \times N_\lambda}$: Observational uncertainty.   
    \item $\mask_{N_{\mathrm {SN}} \times N_{\mathrm {obs}} \times N_\lambda}$: Observational mask, equal to 1 where spectra are valid.    
    \item $\pp_{N_{\mathrm {SN}} \times N_{\mathrm {obs}}}$: Observation time of each spectrum relative to the peak brightness of the supernovae.
    \item $\zz_{N_{\mathrm {SN}} \times N_{\mathrm {latent}} }$: Autoencoder latent space ($\Delta p, \Delta M, \Delta A_V, z_1, ..., z_n$).
    \begin{itemize}
        \item $\Delta p$: Difference in time of peak brightness relative to the SALT2 fits (any float value).
        \item $\Delta M$: Magnitude residual.
        \item $\Delta A_V$: Relative extrinsic extinction.
    \end{itemize}
    \item $\uu_{N_{\mathrm {SN}} \times N_{\mathrm {latent} - 2}}$: Normalizing flow latent space. $N_{\mathrm {latent} -2}$ results from the removal of $\Delta p$ and $\Delta M$.
\end{itemize}
 For purely practical purposes when the number of observations of a supernova is less than the maximum in the data set, N$_{obs}$=32, the end of the data arrays are zero-padded and masked appropriately to fill any remaining empty observation vectors. This has no effect on the model beyond simplifying the training procedure. 

\subsection{Conditional Autoencoder}
\label{sec:AE}
An autoencoder consists of an encoder that maps the input data to a lower dimensional latent representation, and a decoder that reconstructs the data from the latent representation. Both of these components are commonly parameterized as deep neural networks, with weights and biases trained through backpropagation to minimize a loss function. Autoencoders are commonly used for dimensionality reduction or feature learning from an arbitrary dimensional data space, but here we introduce a number of modifications from a general architecture to describe the physical nature of the dataset and separate external and internal information of the supernovae. 

The encoder $\fx$ maps the spectral time series to a latent space $(\Delta p, \Delta M, \Delta A_V, \zz)=\fx(\xx, \pp)$ through a series of fully connected layers of a neural network. Operations are performed along the wavelength axis only, and each spectrum from a supernova is treated independently until the final network layer. The final layer reduces to the mean latent representation of a supernova along the time axis for all observations of the SN~Ia, such that the latent space coordinates of each supernova are forced to represent a compression of the supernova as an object and not a compression of each individual spectra. In this way the entire spectral time series of a supernova is reduced to a few latent variables that together represent a non-linear combination of components describing object-to-object variability, such as the luminosity, spectral tilt, or absorption lines.

Given a latent representation and observation times the decoder $\gz$ learns to reconstruct the spectral time series $\xr=\gz(\fx(\xx, \pp),\pp)$. The decoder first duplicates and concatenates the latent representation $\zz$ with the observation phases $\pp$, then passes this through a number of fully connected layers. Again, the latent and time variable concatenation operations are performed along the wavelength axis only, and each spectrum from a supernova is treated independently. We parameterize the decoder similar to equation~\ref{eq:salt} by not passing some latent parameters through the fully connected layers of the decoder, and instead separating them out to represent an overall amplitude and a phase-independent color law term which are multiplied with the output of the final layer. The encoder and decoder are both fully connected deep neural networks, trained to maximize the agreement of the reconstruction with the data, determined through a loss function $\Loss_{AE}(\xx, \xr)$.

While separating out certain latent parameters to represent physical variables is uncommon in autoencoder literature, it is desirable here due to the physical nature of the problem we are attempting to solve. The time-independent color law term is separated out as we expect that a portion of the color is from the reddenning of the spectrum as it propagates through dust in the intra-galactic medium. This effect is independent of the intrinsic SN~Ia explosion, and so should not propagate through to all the latent variables of the model. The amplitude is separated out as a number of physical effects unrelated to the true cosmological distance can shift the spectra in a way uncorrelated with any spectral features. Peculiar velocity contributions to the redshifts result in an amplitude shift of less than $\sim$10\% for higher redshift SN (z$>$0.02). Additionally the spectra have instrumental ``gray'' offsets of a few percent \cite[e.g.,][]{rubin2022} that also look like an amplitude. These are unique to each spectrum individually, and are typically around $\sim$2\%, but the distribution is very non-Gaussian. Without allowing for an explicit amplitude term in the model any amplitude that is by definition uncorrelated with the spectral features will propagate through to uncertainty in the inferred cosmological distance. Our model thus takes the form of
\begin{align}
\begin{split}
    \label{eq:AE}
    F_{SN}(\pp, \lambda) &= \gz([z_1, ..., z_{n-2}],\pp) \\
    & \hspace{0.5cm} \times 10^{-0.4 [\mathrm{CL}(\lambda) \Delta A_V + \Delta M]},
\end{split}
\end{align}
where $\mathrm{CL}(\lambda)$ can be fit during training by using a single dense layer or can be adopted from physical measurements (e.g., \citep{F99}; we set $R_V=2.8$). We refer to both the $10^{-0.4 \Delta M}$ component and the $\Delta M$ parameter as the extrinsic amplitude of the model throughout this work, but we note that an extrinsic interpretation of $\Delta M$ is degenerate with an intrinsic component that does not vary with wavelength or phase.  

We have found that separating out the amplitude and color law, achieved by physically parameterized decoder architecture shown in Figure~\ref{fig:AE}, does not decrease reconstruction quality compared to a non-physically parameterized autoencoder. In order to match common convention in the literature we rewrite the first three latent parameters as $\Delta p, \Delta M, \Delta A_V$, respectively, and will refer to them as their physical parameter counterparts henceforth. $\Delta p$ represents the difference of the time of peak brightness relative to the SALT2 fits, and while we use the SALT2 time values as an initial guess at the true time of peak brightness, the encoder is free to learn any corrections. As we explain further in the training section, we normalize the average $\Delta p$ over the supernovae to be zero. 

This physical parameterization works to isolate color-like effects into the relative extinction parameter $\Delta A_V$, but there still remains a degeneracy between the relative extinction and output of the decoder determined by the intrinsic latent parameters $z_1, ..., z_n$. While it is possible that a change in $\Delta A_V$ can be counteracted by changing the latent parameters, and thus $\Delta A_V$ is not a direct measurement of the extinction, but rather a measure of the relative extinction between any two supernovae with the same intrinsic latent coordinates.


We chose to use a conditional autoencoder architecture over other time sequence embedding methods due to non-uniformity of the time-step between each observation for each supernova. Long short-term memory networks (LSTMs) \citep{LSTM} are common for sequence-to-sequence predictions, and LSTM Autoencoders are a class used to encode sequence data for a number of applications \citep{LSTM-AE1, LSTM-AE2}, but rely on either time-independent sequential inputs (such as words in a sentence, where one word follows the next with no specified time in between) or on constant time steps between each item of the sequence (such as frames in a video or daily stock prices), and also do not account for missing data. For the purpose of supernovae spectral time series embeddings we have both missing data, such as some supernovae missing observations near peak brightness, and non-uniform time sampling. Some supernovae have a large number of observations spanning the entire ($-10$, $+40$) day time period each separated by $\sim$1-2 days, while others have a small number of observations at more irregular times. For example, our data set has one SN with only 4 observations at ($-5.25, -5.24, +4.65, +14.69$) days. A number of extensions to recurrent models have attempted to deal with missing timesteps through masking \citep{RNN-missing-time}, by incorporating the passing of time in between observations in a ``time-aware LSTM'' through weighting the short term memory by the elapsed time \citep{LSTM-time-aware}, or a ``Phased LSTM'' \citep{LSTM-phased} that adds a new oscillating time gate which only updates the networks weights during a small percentage of the cycle. While these have potential to work for our application, the small amount of training data available (relative to standard machine learning benchmark data sets) and non-uniform sampling, coupled with the desire for a physically parameterized interpretable network for posterior analysis, led us to stick with a more standard autoencoder setup, although initial investigations using a LSTM autoencoder did not perform poorly.

\subsubsection{Normalizing Flow}
\label{sec:NF}
Once the autoencoder is trained its parameters are fixed and we determine the latent space prior $P(\zz)$ probabilistically by constructing a bijective mapping $\bz$ from the latent space $\zz$ to a Gaussian latent space $\uu = \bz(\zz)$. A forward pass of the bijective mapping ($\zz \rightarrow \uu$) allows for rapid density estimation of a point in the $\zz$ space, while an inverse pass ($\uu \rightarrow \zz$) allows for sampling of the $\zz$ space. We determine this mapping through a normalizing flow (NF), popularized by \citet{realnvp, MAF} and comprehensively reviewed in \citet{normalizing_flows}. The NF is parameterized by a fully connected deep neural network and trained to minimize the negative log likelihood of the encoded samples, where the NF prior is a unit Gaussian, $\bm{p}(\uu) = \NN(0,1)$. {\textit{To ensure the model is not dependent on cosmological parameters, peculiar velocities, or gray offsets, we do not include the amplitude term $\Delta M$ in the normalizing flow}}. In this fashion we do not impose any prior for the amplitude. 

With both a trained autoencoder and normalizing flow we have a fully probabilistic and generative model capable of generating new samples $\xp$ from the data distribution $\bm{p}(\xx)$ as follows (illustrated in Figure~\ref{fig:AE}):
\begin{enumerate}
    \item Draw a sample $\uu$ from $\NN(\bm{0},\bm{1})$
    \item Pass the sample through the normalizing flow to get $\zz_u = \bz^{-1}(\uu)$
    \item Concatenate $\zz_u$ with the desired amplitude offset $\Delta M$ to get $\zz$.
    \item Pass $\zz$ and the desired observation times $\pp$ through the decoder, $\xp=\gz(\zz,\pp)$. Empty observation timeslots are automatically masked appropriately.
\end{enumerate}

\subsection{Posterior Analysis for uncertainty quantification}
\label{sec:posterior_methods}

After training is completed, the PAE can be used to provide uncertainty quantification on the best-fit latent parameters of the model of equation~\ref{eq:AE} ($\Delta p$, $\Delta M$, $\Delta A_V$, $z_1$, ..., $z_n$). The log posterior of a data point under the PAE is \citep{UQ}
\begin{align}
\begin{split}
    \label{eq:likelihood}
    \mathrm{ln}\ P(\uu | \xx, \pp)  &= \mathrm{ln}\ P(\xx | \uu, \pp, \sig_{\mathrm{noise}})/N_{\mathrm{spectra},SN} \\
    & \hspace{0.5cm} + \mathrm{ln}\ P(\uu) \\
    & \hspace{0.5cm} + \mathrm{const},
\end{split}
\end{align}
where the prior is $P(\uu)=\NN(0,1)$,  and the implicit likelihood is given by $P(\xx| \uu, \pp, \sig_{\mathrm{noise}}) = \NN(\gz(\bz^{-1}(\uu),\pp), \sig^2 + \sig_{\mathrm{recon}}(\pp)^2)$. Note that we replace the data $\xx$ by its generative process $\gz(\bz^{-1}(\uu),\pp)$, which brings the inference problem to the low dimensional latent space of the PAE, making the posterior analysis much more computationally tractable. 

The covariance of the Gaussian likelihood has two terms: the PAE reconstruction error $\sig_{\mathrm{recon}}(\pp)$, and the noise level in the data $\sig$. We measure the PAE reconstruction error as a function of observation time by binning the test data in 5 day intervals and linearly interpolating when performing the posterior analysis. The model uncertainty is calculated as a fraction of the observed flux rather than the standard deviation.  

For each supernova we perform latent space posterior analysis in order to find the best-fit data reconstruction under the PAE model. In addition to the intrinsic latent parameters included in the normalizing flow, we have a free time shift parameter $\Delta p$ to allow for a different time-origin $p=0$ than the SALT2 fits used for initial model training, and the extrinsic magnitude residual $\Delta M$. Therefore, the posterior analysis takes the form of 
\begin{equation}
    \xr_{\mathrm{recon}} = \gz([\Delta M, \bz^{-1}(\uu')],\pp+\Delta p),
\end{equation}
where we simultaneously fit for the $\Delta M$, $\uu'$, and $\Delta p$ values which best reconstruct the spectral time series for each supernova. A Gaussian prior on the time shift can be added, as the uncertainty on the time of the peak luminosity is generally half a day \citep{snemo}, but we found that unnecessary here.

To find the maximum of the posterior (MAP) we begin optimization from the best fit encoded value of the data, as well as 24 additional initialized points in the ($\Delta M$, $\uu$, $\Delta p$) parameter space. Optimization is performed using the Limited-memory BFGS (LBFGS) algorithm. To ensure that we converge to the global minimum, and not some local minimum near $\Delta M = 0$, we sample these 24 initialization points from a much larger region than the prior distribution $\uu$ and thus the variations of any parameter are not artificially small due to any limitations of the encoder. We initialize 10 points with a magnitude residual linearly spaced between $\Delta M = (-1.0, 1.0)$, and 10 points linearly spaced between $\Delta A_v = (-0.5, 3.0)$. Sampling these large magnitude residual and extinction values ensures that supernovae with high velocities or levels of dust will still have an initialization value nearby, and that we will probe the true minimum of the posterior. We find minimal spread of the minima found from the 25 initialization points for the majority of supernovae, except for the most nearby SNe~Ia, whose peculiar velocities can be a large fraction of the total redshift, for which the large spread in initializations is required to converge to the best fit value.     

We denote the MAP latent variables as the best fit parameters that maximize the posterior from these 25 minima. From the MAP value we then run Hamiltonian Monte Carlo (HMC) \citep{HMC} - a Markov chain Monte Carlo (MCMC) algorithm that takes a series of gradient-informed steps to produce a Metropolis proposal - to marginalize over ($\Delta M$, $\uu$, $\Delta p$) to obtain the final best fit model parameters and their uncertainty. While this procedure provides best fit parameters nearly equivalent to the MAP values, it provides a more robust estimation of their uncertainty. HMC is run for 25,000 iterations following a 10,000 step burn-in in which the step size is allowed to vary to target an acceptance rate of 0.651 \citep{HMC_tuning}. 

\section{Architecture \& Training}
\label{sec:architecture}
The unique PAE architecture coupled with the physical nature of the data required a number of modifications compared to the training of a standard autoencoder, including a multi-stage training setup and significant data augmentations. Models are trained in Tensorflow \citep{tensorflow}, utilizing Tensorflow Probability \citep{tensorflow_probability}.

\subsection{Limited Dataset}
\label{sec:regularization}
A severe limitation to training deep neural network architectures on a SN~Ia dataset, compared to typical machine learning datasets, is the very limited number of data samples (only 228 supernovae in this work). Therefore to prevent overfitting we implemented a number of techniques throughout training, including early-stopping, dropout, data augmentation, and weight regularization. 
\begin{itemize}
    \item Dropout: As each supernova provides a time series, we experimented with two types of dropout. The first is as usual, dropping out neurons in the encoder with dropout rate=0.2, ensuring that the dropout mask is the same for all timesteps. The second is dropping out a random sample of 10\% of the spectra for each supernova at each training epoch. This was chosen to negate the small number of spectra that seem to have higher measurement error, and/or do not follow as consistent of time-series evolution with neighbouring observations. This spectral dropout helped training the most, while standard dropout had limited success, likely due to the small size of the training data available.
    \item L$_\mathrm{2}$ regularization: we implemented L$_\mathrm{2}$ regularization. L$_\mathrm{1}$ kernel regularization was not appropriate for this problem, as we do not want to encourage sparsity in our model.
    \item Data Augmentation: At each epoch a random Gaussian noise draw consistent with the observational error was added to each spectrum, $\xx_{\mathrm{epoch}} = \xx + \NN(0,\sigma_{\mathrm{noise}})$. At each epoch we also vary the SALT2 phase given for each SN by a random Gaussian draw from $\NN(0, \sigma_{t,\mathrm{SN}})$, where $\sigma_{t,\mathrm{SN}}$ is the measured value from the SALT2 light curve fits and is unique for each supernova.
    \item Early stopping: We experimented evaluating model performance on a validation set every 100 epochs, and during early trials experimented with keeping the model that achieved the lowest reconstruction loss. We found this had a negligible effect over keeping the final epoch, as the amount of regularization employed and the small latent space was sufficient to prevent the model from overfitting. Therefore we do not employ early stopping, do not use a validation set in addition to the training set, and the test data is unseen until model evaluation.
\end{itemize}

\subsection{Loss function}

The loss function we used has two terms. The first is a standard reconstruction error, while the second is a latent space correlation penalty term to discourage correlations between any desired terms of the model.

For the reconstruction error term we investigated a number of loss functions. Unlike many machine learning datasets that only include data samples, we also have the measurement uncertainty for each wavelength bin of each sample. A bad reconstruction of a sample with small measurement errors should be discouraged more than on a sample with large measurement errors, which a standard e.g., mean squared error, loss term does not account for. We investigated a number of loss functions including mean absolute error, mean square error, Huber loss, and the error-weighted counterparts of these, the negative Gaussian log likelihood, and the square root of each of the previous. We found the best reconstructions when using a weighted Huber loss for the reconstruction error term with $\delta=25$:

\begin{equation}
   \Loss_{\mathrm{recon}}(SN) = \sum_{N_{obs}, N_\lambda}
\begin{cases} 
      \frac{1}{2} \left(\frac{\xx - \xr}{\sigma} \right)^2 \mask & \left| \frac{\xx - \xr}{\sig} \right| \leq \delta \\
      \delta \left | \frac{\xx - \xr}{\sigma} \right| \mask - \frac{1}{2}\delta^2 & \left| \frac{\xx - \xr}{\sig} \right| > \delta,
\end{cases}
\end{equation}
where $\sig$ is the measurement uncertainty of the observed spectra and $\mask$ is the mask specifying valid wavelength bins. The Huber loss scales as the mean squared error when the noise-weighted error is smaller than $\delta$, and as the mean absolute error when the noise-weighted error is greater than $\delta$. This loss helps with the few measurements that are very large outliers to the expected spectral envelope.


Each supernova has a different number of observations and thus the arrays were zero padded. Therefore we only compute the loss over the existing number of observations for each, $N_{obs}^{SN}$. This results in supernovae with more observations being given a larger weight in the training process, and is equivalent to weighting the loss by the number of observations per supernovae. 

\begin{figure*}
\centering
\includegraphics[width=0.33\textwidth]{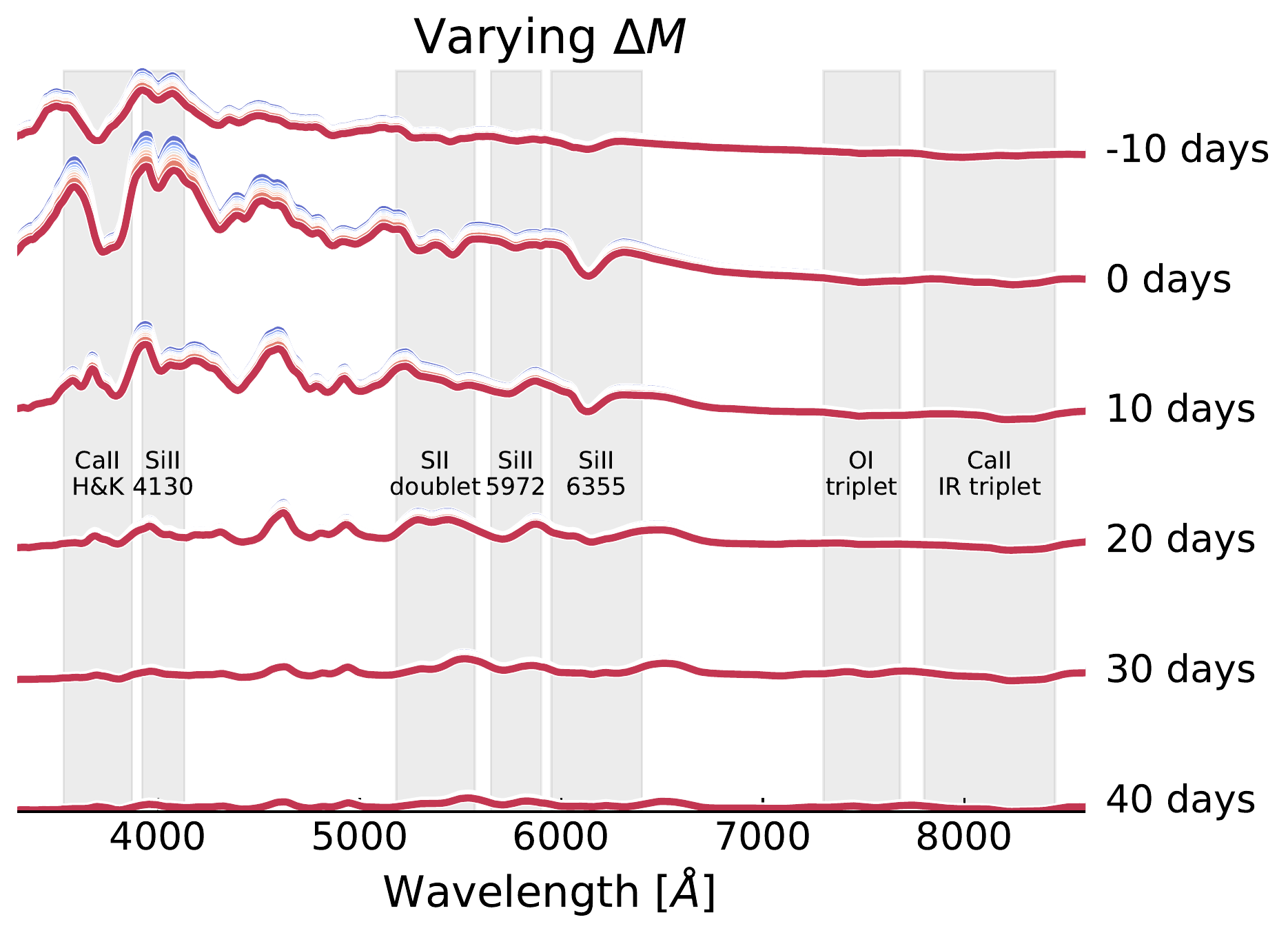}
\includegraphics[width=0.33\textwidth]{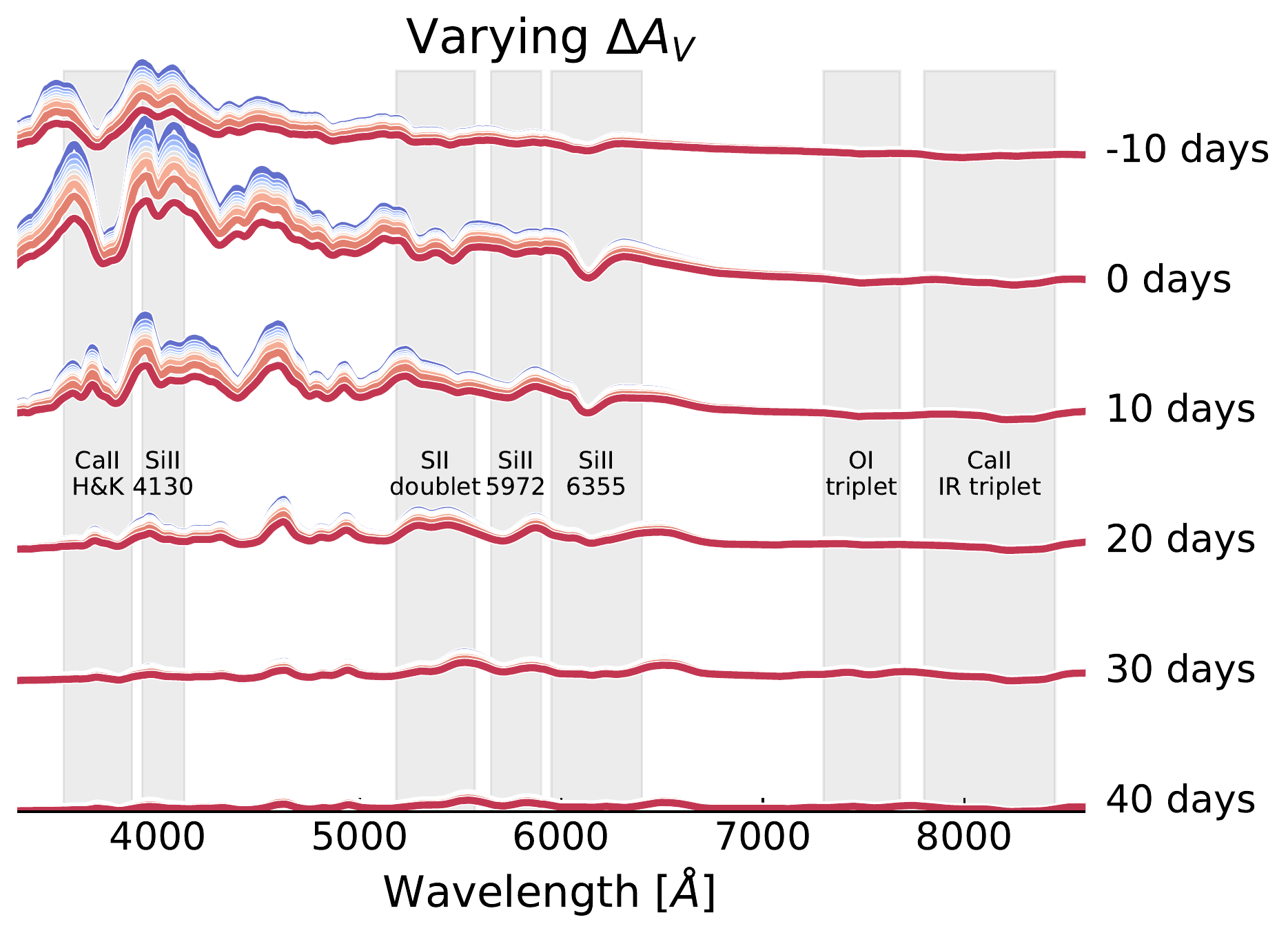}
\includegraphics[width=0.33\textwidth]{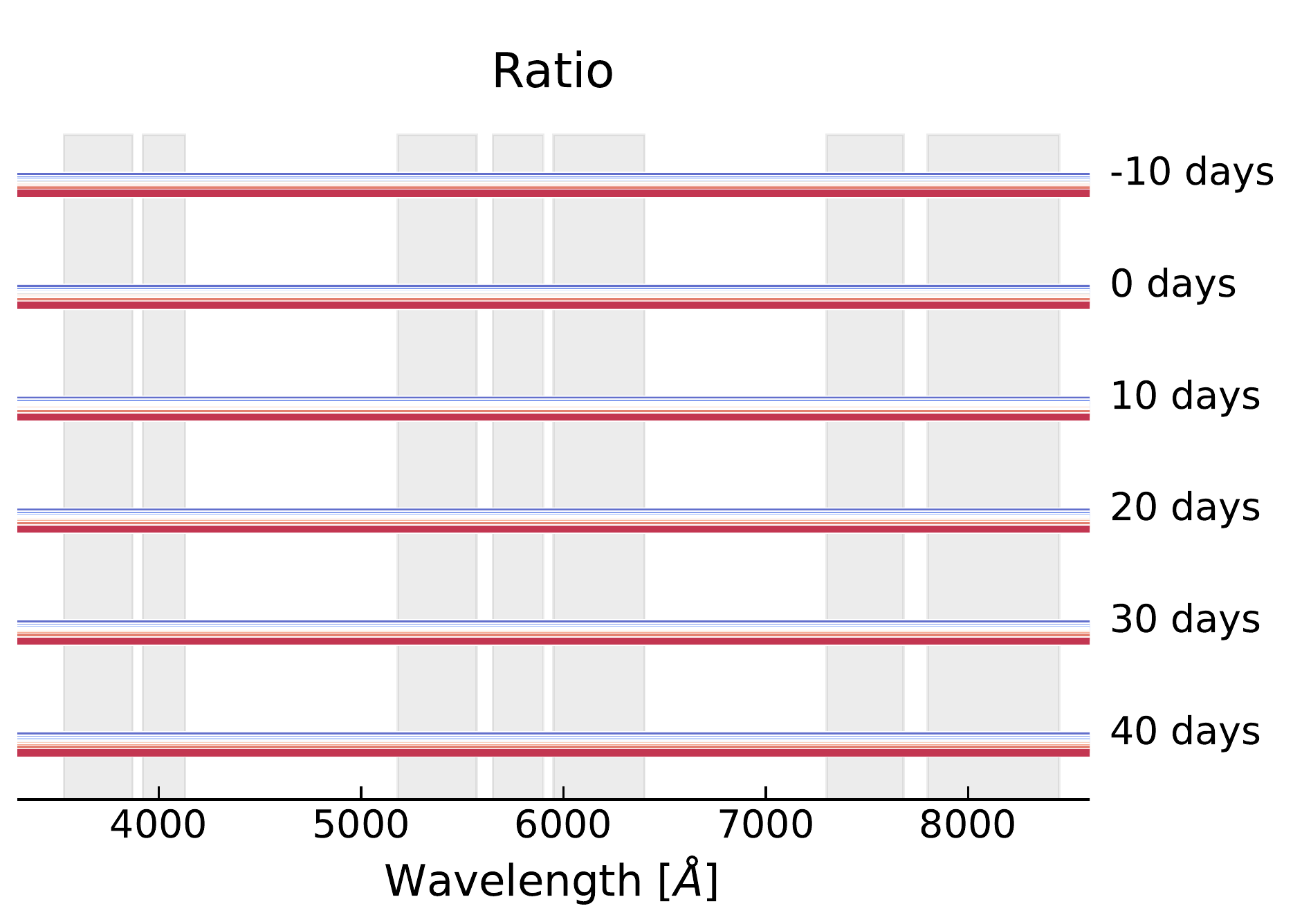}
\includegraphics[width=0.33\textwidth]{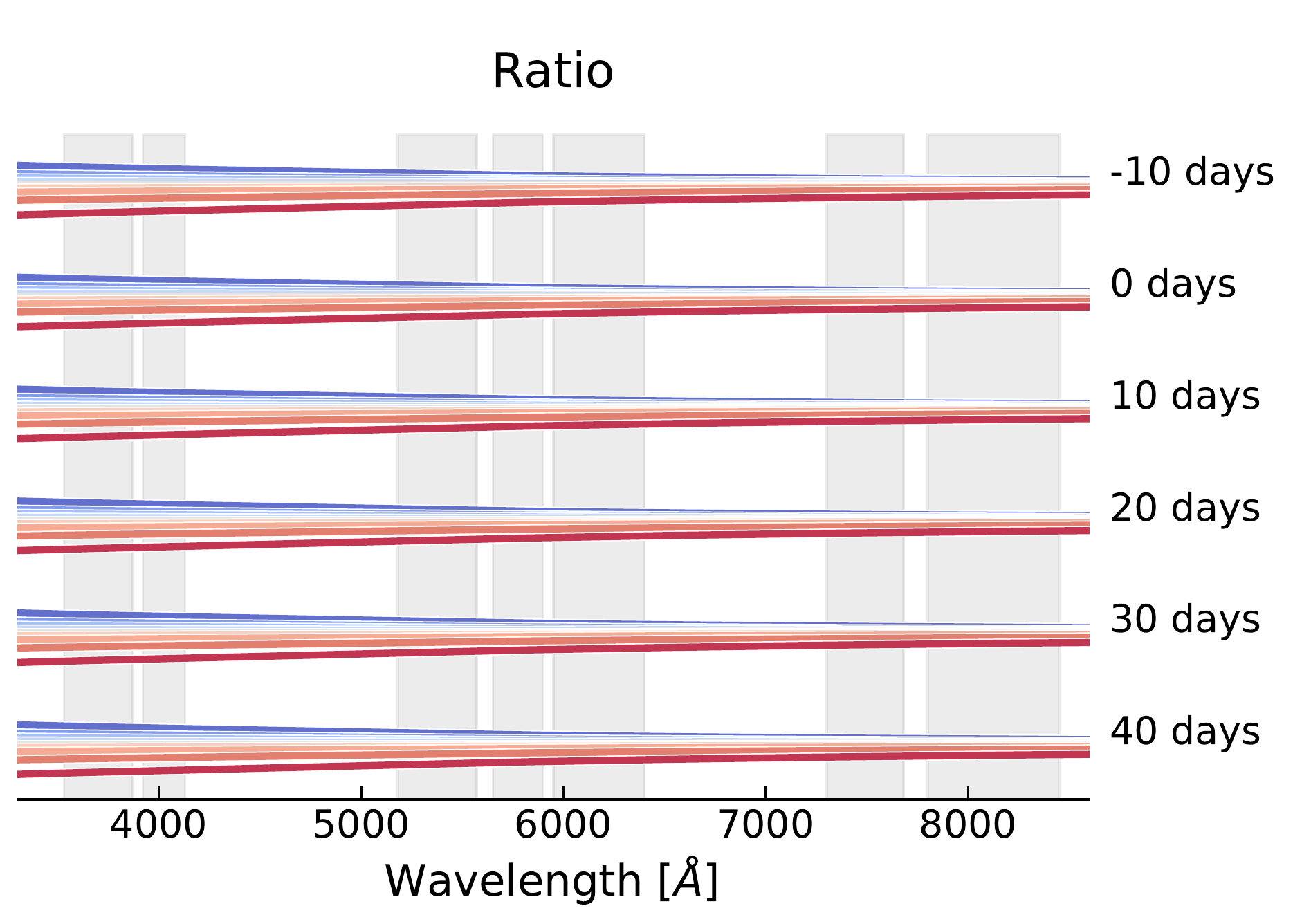}

\includegraphics[width=0.32\textwidth]{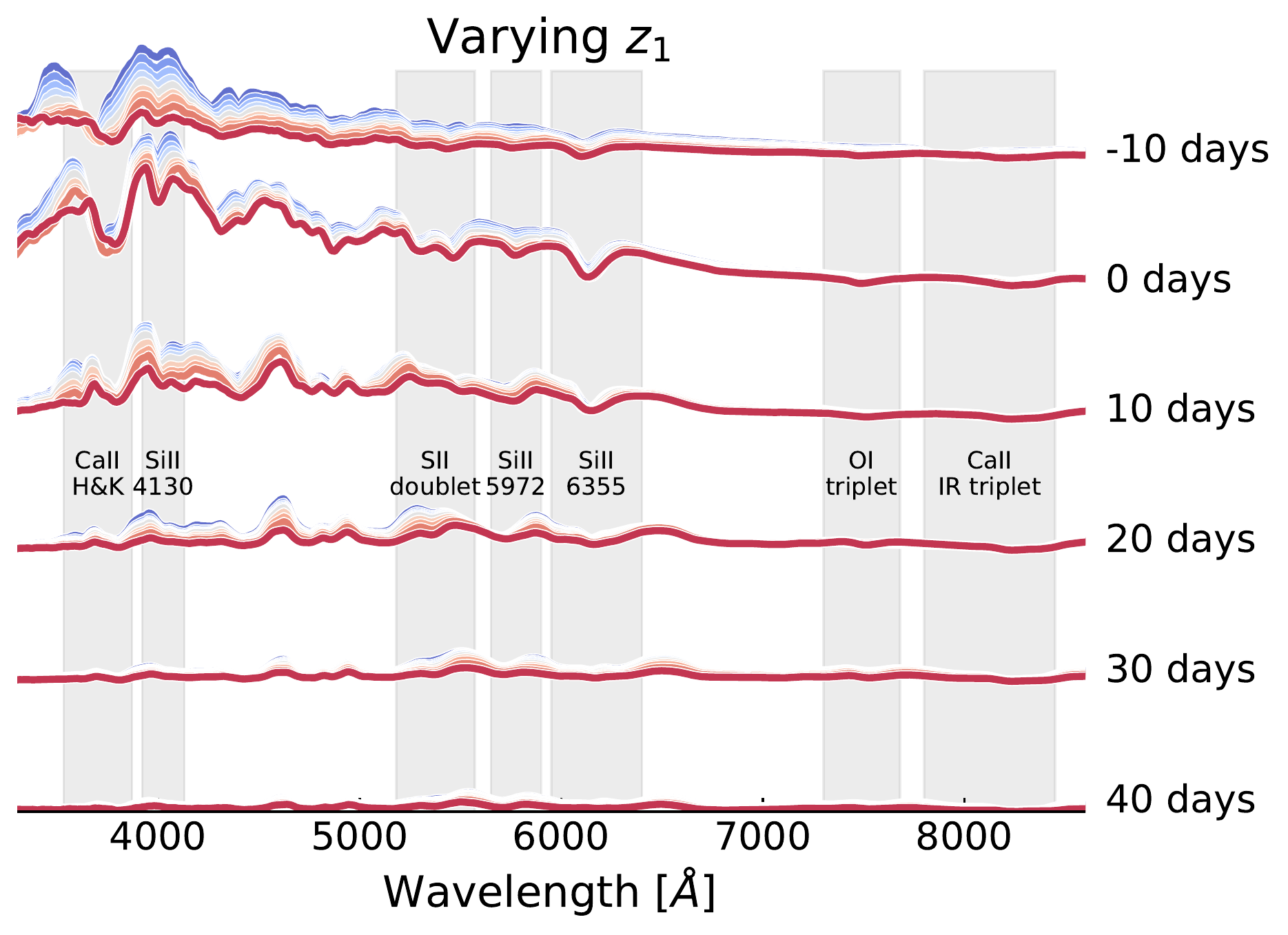}
\includegraphics[width=0.32\textwidth]{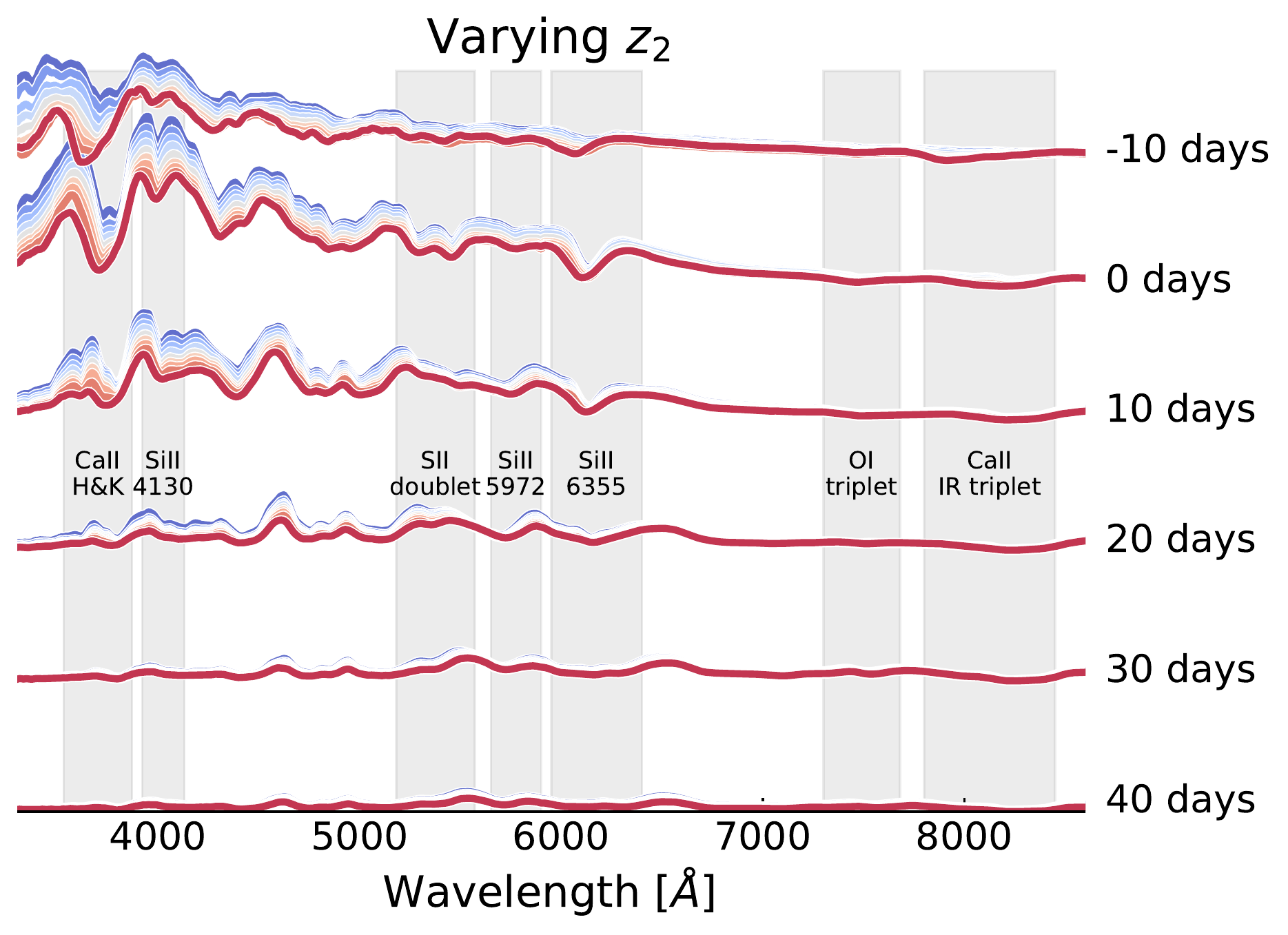}
\includegraphics[width=0.32\textwidth]{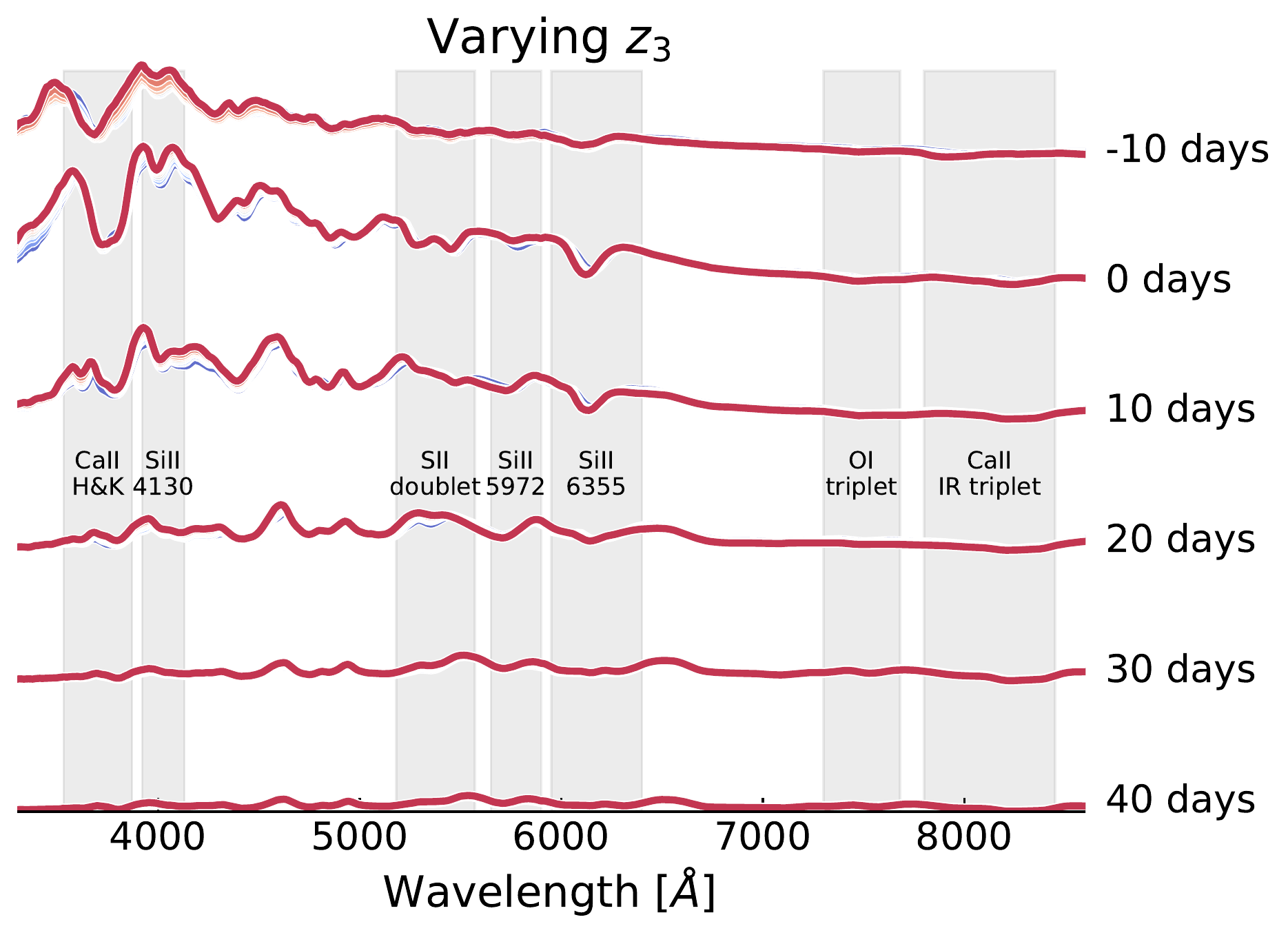}

\includegraphics[width=0.32\textwidth]{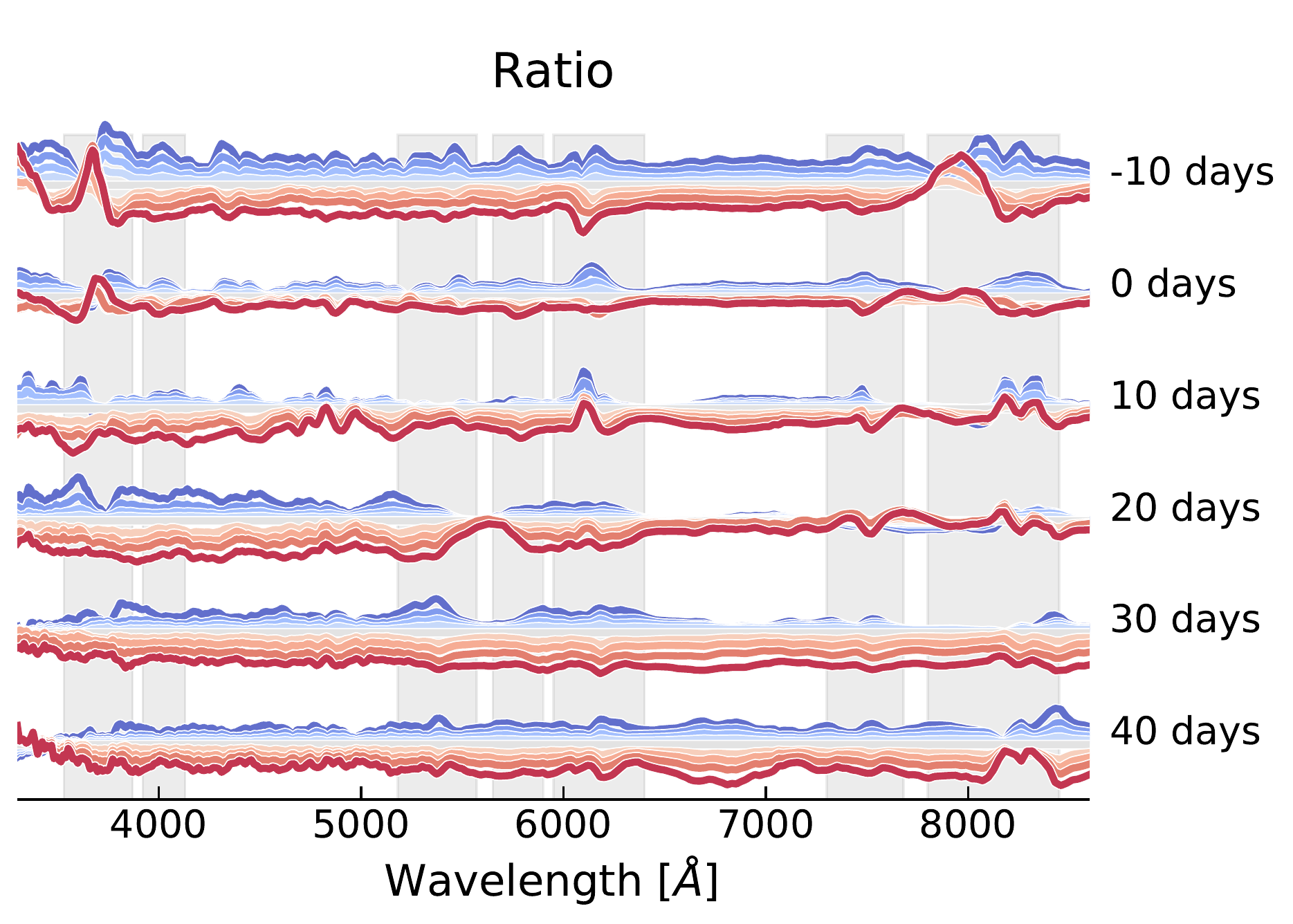}
\includegraphics[width=0.32\textwidth]{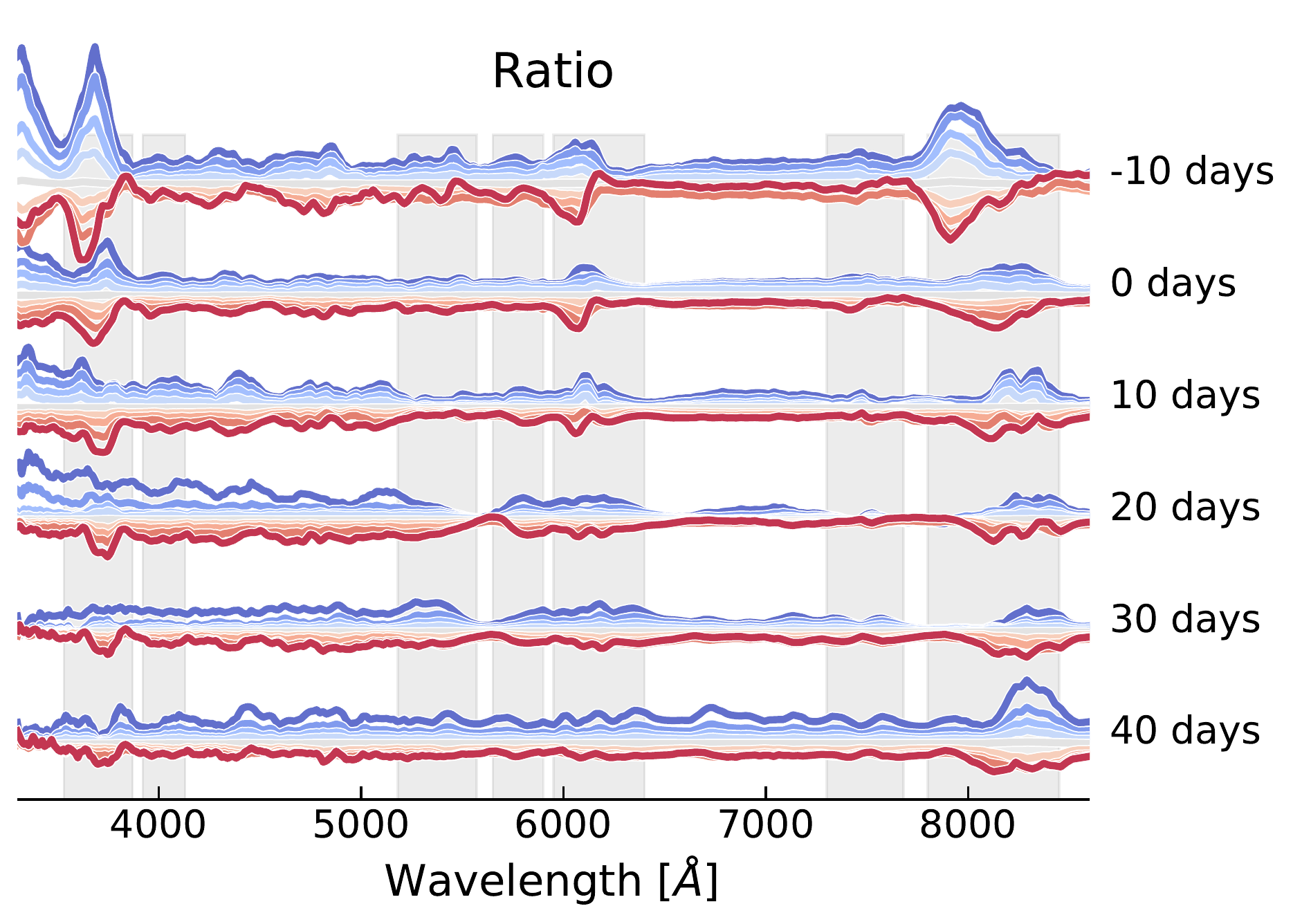}
\includegraphics[width=0.32\textwidth]{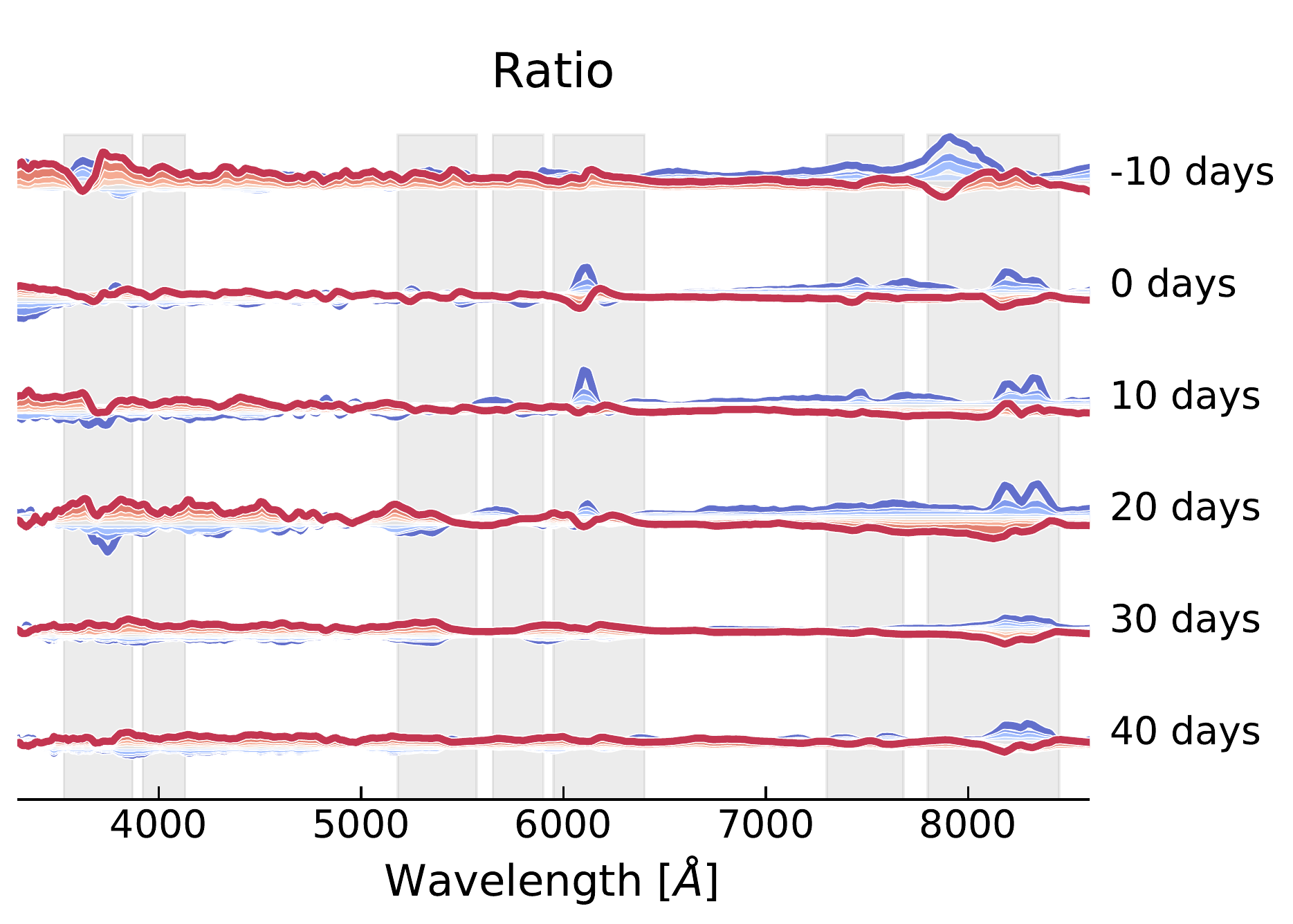}

\includegraphics[width=0.32\textwidth, trim=0 0 -45 0,clip]{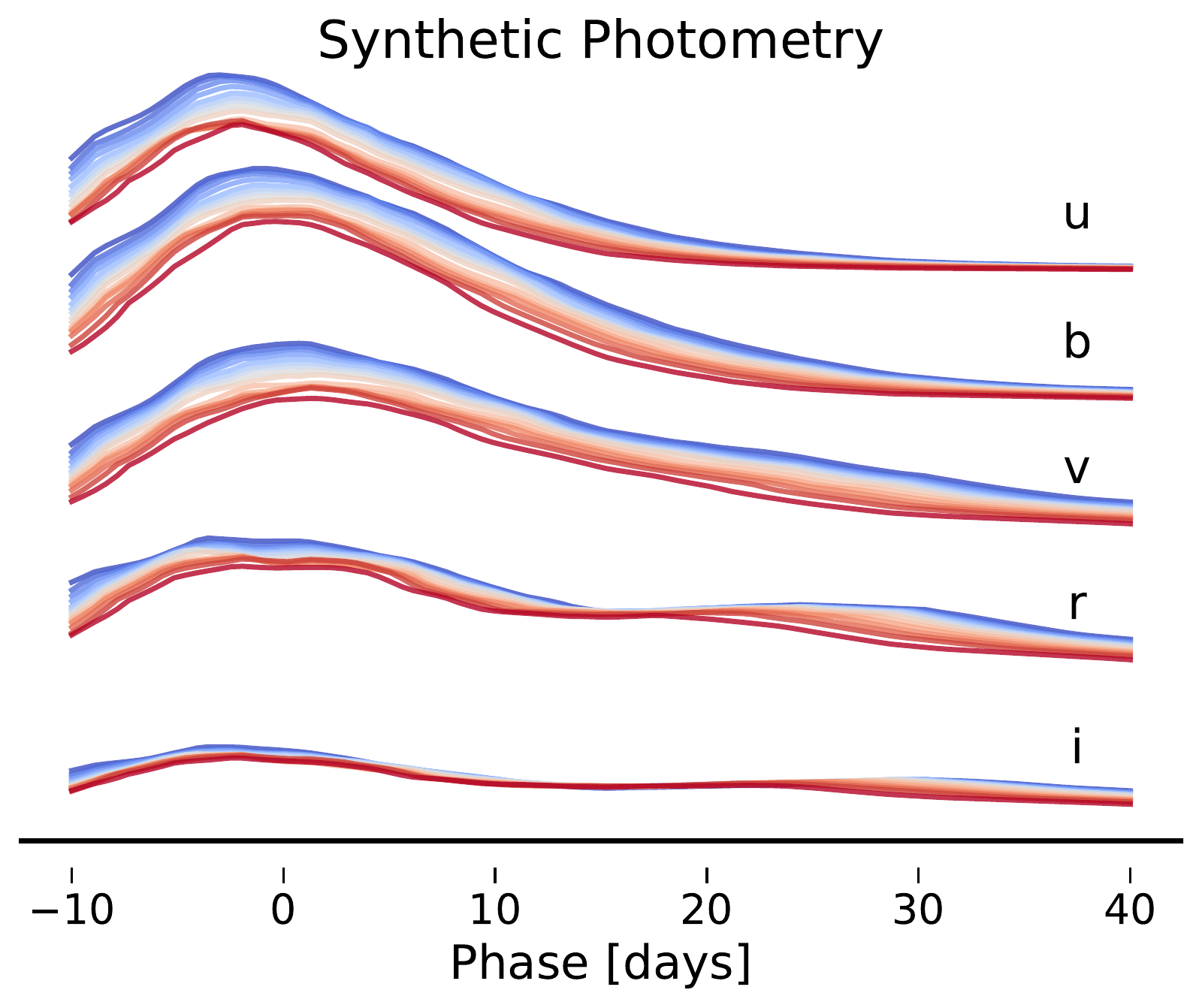}
\includegraphics[width=0.32\textwidth, trim=0 0 -45 0,clip]{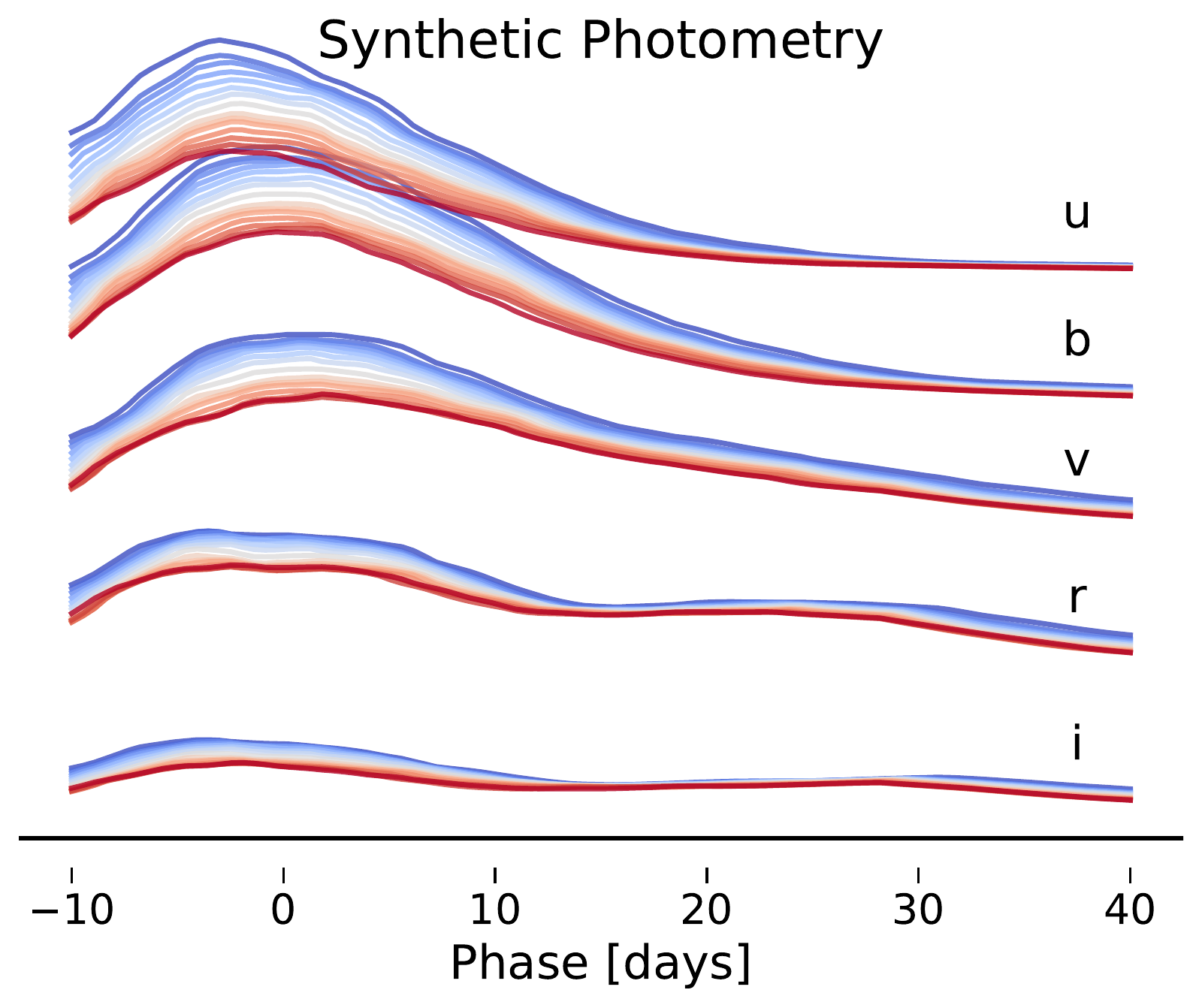}
\includegraphics[width=0.32\textwidth, trim=0 0 -45 0,clip]{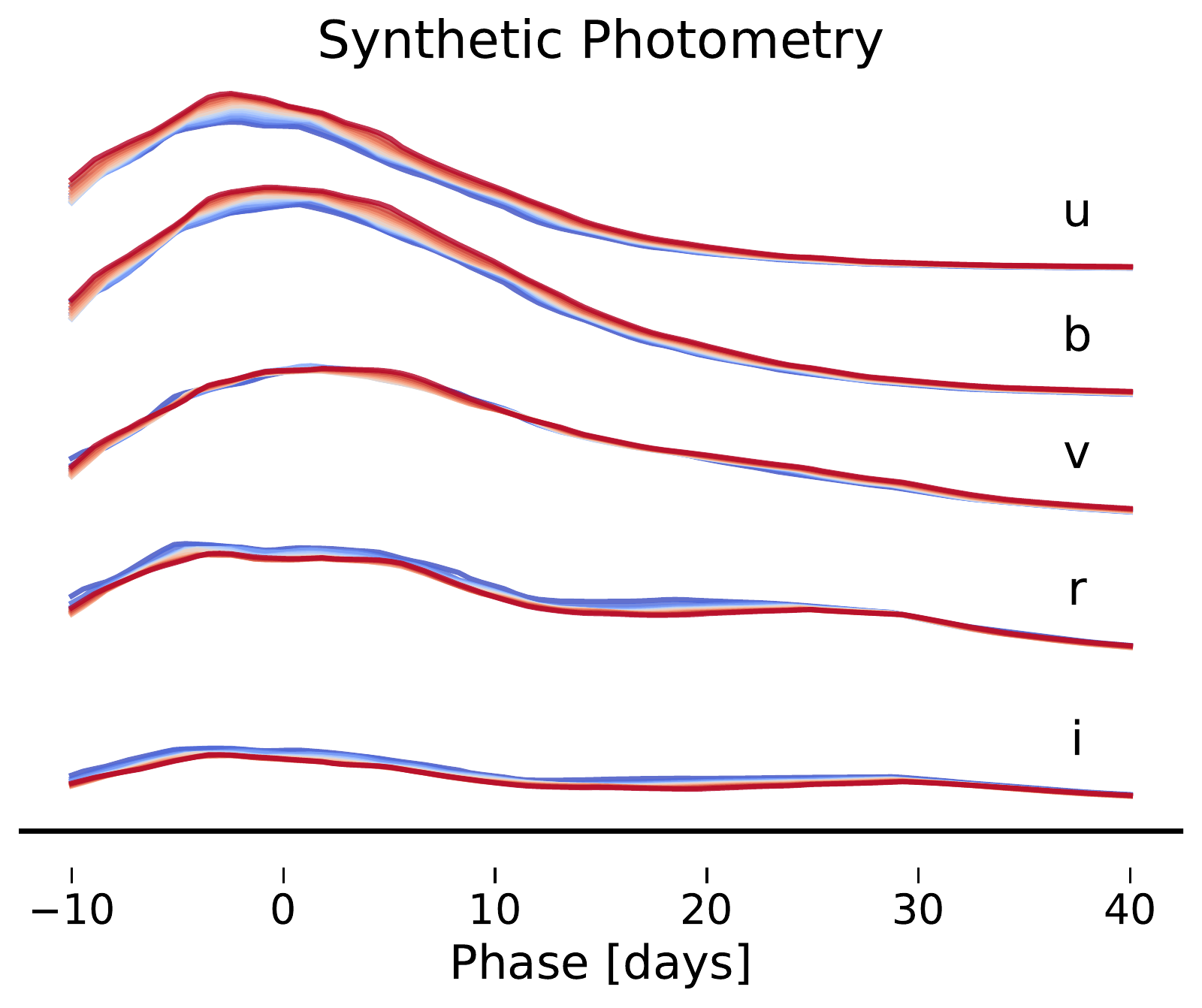}

\caption{Generative sampling of SN~Ia spectra as a function of phase, individually varying each latent dimension of a PAE model with two extrinsic ($\Delta M, \Delta A_V$) and three intrinsic non-linear ($z_1, z_2, z_3$) parameters, while keeping the other latent variables fixed at their mean values. The top panels of each set show the spectra with a constant offset in luminosity, while the bottom show the ratio from the mean. For the non-linear parameters we also display the resulting synthetic photometry. Blue lines are values lower than the mean, transitioning through red for values higher than the mean.}
\label{fig:vary_latent}
\end{figure*}

A key use of the PAE will be to constrain the most likely latent parameters and their uncertainty for each SN~Ia. Specifically, the amplitude $10^{-0.4\Delta M}$ is key to constrain the intrinsic luminosity of the SN, and therefore can be used to estimate the distance and distance uncertainty to the object. When unconstrained during training, this parameter will learn both the intrinsic diversity that affects the spectrum in a similar manner to a brightness difference, and the extrinsic diversity from peculiar velocities and gray offsets. We expect that the intrinsic diversity, although similar to an amplitude offset, is correlated with features of the spectra, while the extrinsic diversity is by definition uncorrelated. Both the SALT2 and Twins Embedding models let the amplitude contain both intrinsic and extrinsic luminosity components, and then implement an additional step to ``standardize'' the magnitude residuals in an attempt to explain the intrinsic luminosity contribution as a linear or non-linear function of the remaining model parameters. 

During training, instead of letting the amplitude freely vary to explain both the intrinsic and extrinsic luminosity and then fitting additional models to explain the two terms, we encourage the latent space to learn latent features that are uncorrelated with the amplitude. {\textit{This ensures that the amplitude term of the PAE directly models only the extrinsic amplitude component}}, and the intrinsic luminosity is described by a non-linear combination of the remaining latent parameters.

To achieve this we added a loss term proportional to the correlation between the amplitude and the other latent parameters. This is a similar idea to \citet{PCAAE}, who additionally went further to learn an organized latent space with their ``PCA Autoencoder'', which learns each latent dimension using a separate autoencoder in a series of encoder-decoder pairs. Here we include a latent space correlation coefficient penalty,
\begin{equation}
    \Loss_{\mathrm{corr}} = \sum_{z_i, z_j}  \left(  \frac{ {\mathrm{Cov}}(z_i, z_j) }{ \sqrt{ \mathrm{var}(z_i) \mathrm{var}(z_j) }} \right)^2 \times \mathrm{Mask}, 
\end{equation}
where the mask can allow for correlations between intrinsic latent parameters ($=0$ on the diagonal and block of $z_1, ..., z_n$ terms, and $=1$ otherwise), or can discourage correlations between any parameters ($=0$ on the diagonal and $=1$ otherwise).



The total loss function that the autoencoder is trained on then becomes
\begin{equation}
    \Loss_{\mathrm{AE}} = \Loss_{\mathrm{recon}} +
    \lambda_{\mathrm{corr}} \Loss_{\mathrm{corr}},
\end{equation}
where $\lambda_{corr}$ is a free parameter whose value we chose to return similar  values from the reconstruction and correlation loss terms in the early stages of model training. We found that this correlation penalty helped to uncorrelate the latent parameters with nearly no reduction of reconstruction accuracy.

\subsection{Training}
\label{sec:training}
A number of architectures and training methods have been investigated for the autoencoder. We found that a fully connected architecture performed better than a convolutional one, and found the lowest reconstruction error when using 3 encoding and decoding hidden layers with (256, 128, 32) and (32, 128, 256) neurons in each layer, respectively. We found best performance when using rectified linear activations (ReLU) \citep{relu1, relu2} for each hidden layer, and no activation on the final output of the encoder or decoder. This results in a large amount of trainable parameters -- nearly 112,000 in each of the encoder and decoder models. Compared to the number of spectra used (2696) with 288 spectral bins each (for a total of 776,448 degrees of freedom), the number of model parameters is sizeable, but in practice it has been shown that heavily parameterized neural networks empirically improve both optimization and generalization \cite{zhang_generalization, overparameterizedNN}, while allowing the model to represent much more complicated functions than ones with fewer parameters. During model training we employ a number of regularization methods (discussed in Section~\ref{sec:regularization}), and find no evidence for overfitting.  

We trained the autoencoder in four separate stages, using ADAMW learning rate optimization \citep{ADAMW}. In the first stage, we set the extrinsic magnitude dispersion $\Delta M$, the time difference relative to the SALT2 fits $\Delta p$, and the non-physical latent parameters $z_i$, to zero while letting the relative extinction $\Delta A_V$ vary over 1000 training epochs. In the second stage, we initialize the encoder and decoder weights and biases with the values learned from the first stage, randomly reinitialize the weights of the final layer corresponding to the non-$\Delta A_V$ parameters (using TensorFlow's GlorotUniform initializer, and scaling the weights down by a factor of 100), and again train both the encoder and decoder while now also allowing the non-physical latent parameters to freely vary, for 1000 epochs. The third step is analogous, now also allowing $\Delta M$ to vary, and we train for 5000 epochs. The final stage lowers the learning rate from the 0.005 used in the previous steps to 0.001, and also allows $\Delta p$ to vary. Each training stage employs weight decay regularization with an initial value of 0.0001, and both the learning rate and weight decay factor follow an exponential decay scheduler with a decay rate of 0.95 over 300 steps.

We found that this multi-stage learning procedure helped to utilize the physical parameters of the model, specifically the relative extinction $\Delta A_v$, which otherwise often got stuck in local minima near its initialized value. Separating the first two training stages significantly decreased the level of intrinsic amplitude that remained in $\Delta M$, even when utilizing the correlation penalty. We also found that learning $\Delta p$ in a separate final stage improved reconstruction accuracy over introducing at the beginning of model training, as by this point the PAE had already learned an accurate description of SNe~Ia evolution and the introduction of $\Delta p$ simply decouples from the initial estimate of the SALT2 model. During training we enforce $\overline{\Delta p} = 0$, $\overline{\Delta M} = 0$ (i.e. mean amplitude $= 1$), and $\overline{\Delta A_V} = 0$, by a custom layer similar to a batch normalization, but only standardizing the mean to zero and not the variance. This is achieved by subtracting the mean $(\Delta p, \Delta M, \Delta A_v)$ of each batch during training from the output of the encoder before feeding the latent parameters to the decoder. When training is complete we calculate the mean over the entire training set, and hardcode the encoder to subtract this mean. These modifications have no effect on the data reconstruction quality, but ensure that the parameters represents the difference from the average supernova. Ensuring $\overline{\Delta p} = 0$ produces a phase that is in sync with the SALT2 fit (on average), but for individual supernova can decouple the PAE phase from the maximum luminosity in B-band.  

We use 75\% of the supernovae for training, and reserve 25\% for testing, for totals of 171 training and 57 testing samples. The latent space correlation penalty motivates a large batch size in order to properly evaluate the correlations between supernovae, so we utilize a batch size of 57. As we do not employ early stopping, our model has not learned using any information from the test set, although when examining the test at a later time we do not find evidence of overfitting -- the reconstruction error on the test set continues to decrease, or flattens, throughout training and does not then begin to increase. While training we impose a minimum redshift cut of $z>0.02$ to negate the significant peculiar velocity contribution to the low redshift samples, such that only 145 of the 171 training samples are used for learning. Spectra amplitudes as given in reference frame units are already scaled to $\sim (0-1)$ so we do not further scale the spectra. We minmax scale the times of observation to range between (0, 1) instead of ($-10$, $+40$) days. Spectra with any masked wavelength bins are not used in the encoder as spurious values will propagate through to the latent variables, but the non-masked portions of the reconstructed spectra are used in the calculation of the loss.

Our normalizing flow to transform from latent variables $\zz$ to latent Gaussian variables $\uu$ is implemented as a Masked Autoregressive Flow (MAF) \citep{MAF}. As stated previously we do not include the $\Delta M$ parameter in the normalizing flow in order to ensure that there is no prior on the extrinsic amplitude. The normalizing flow is not conditional, as the $\zz$ variables do not depend on the observation time. We use 12 layers with 8 units per layer, and train for 500 epochs on the training data using the ADAM optimizer \citep{ADAM}, splitting 33\% off as a validation sample, and stopping when the log probability on the validation sample does not decrease for 30 epochs. This early stopping was required for the flow, as we found it has the potential to overfit. The small size of the flow relative to the decoder means that its computational cost is a negligible fraction of the posterior analysis. As such we did not perform an architecture search to minimize the size of the flow.

\section{Results}
\label{sec:results}

\begin{figure*}
\centering
\includegraphics[width=0.49\textwidth]{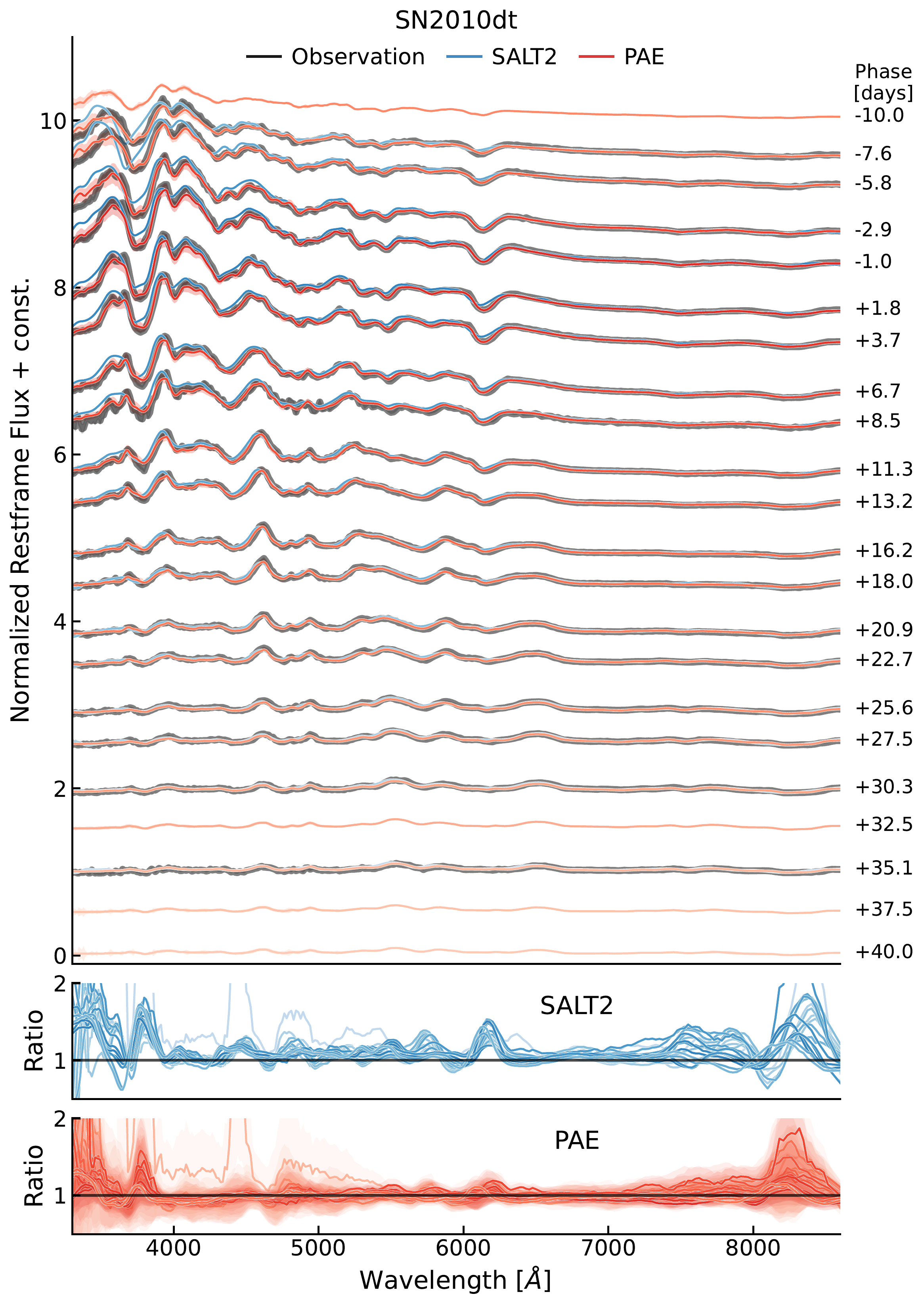}
\includegraphics[width=0.49\textwidth]{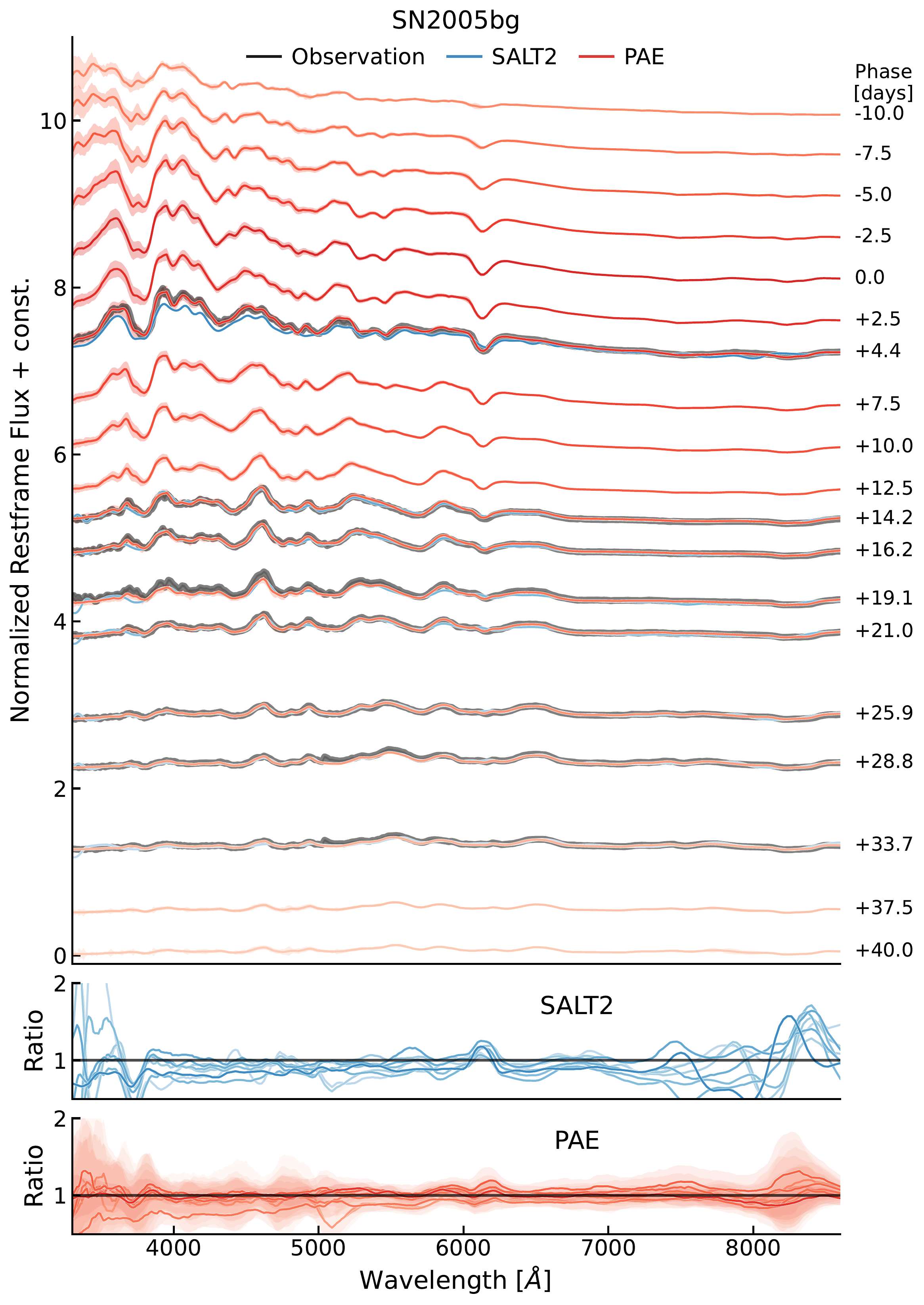}

\includegraphics[width=0.49\textwidth]{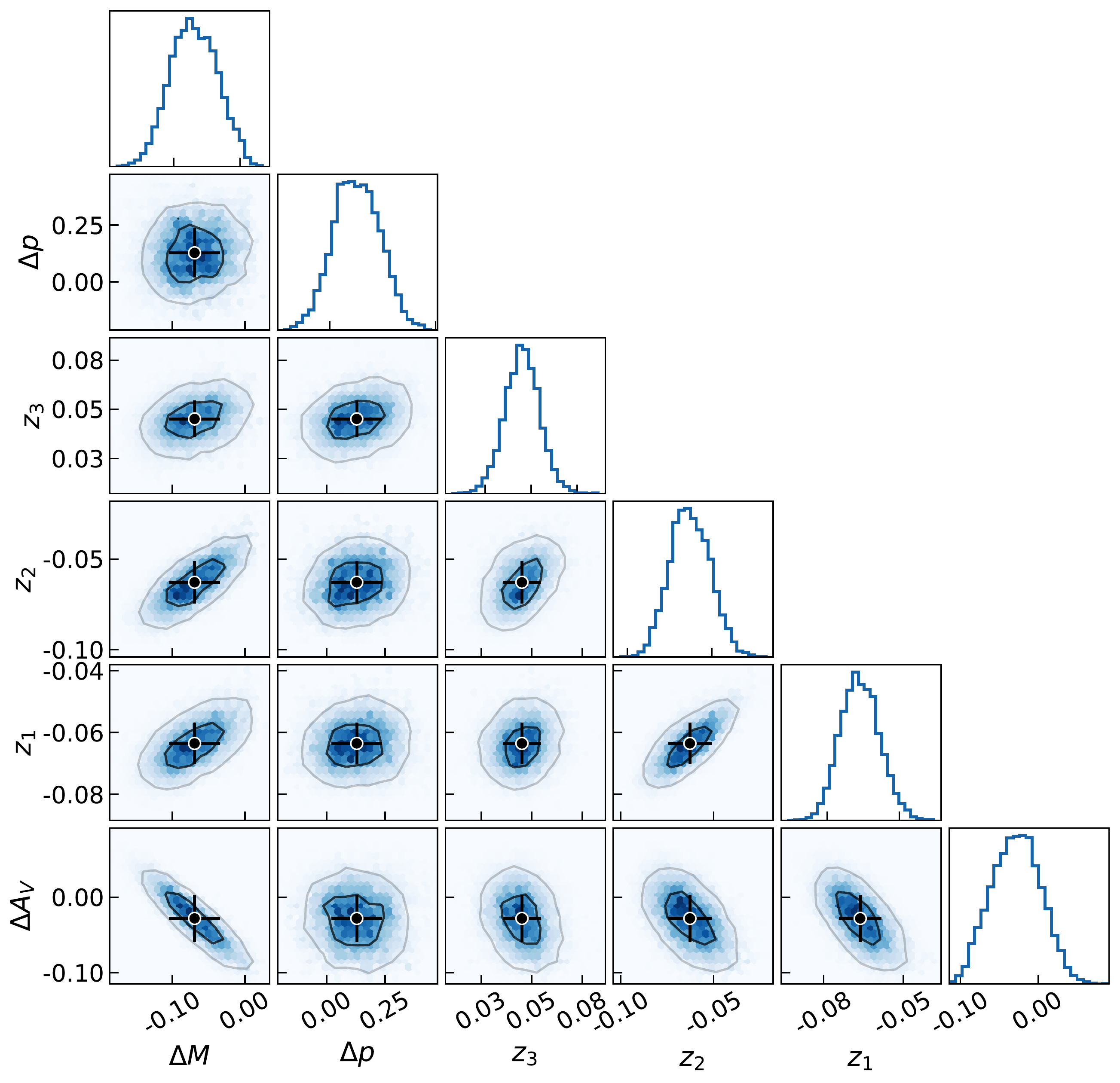}
\includegraphics[width=0.49\textwidth]{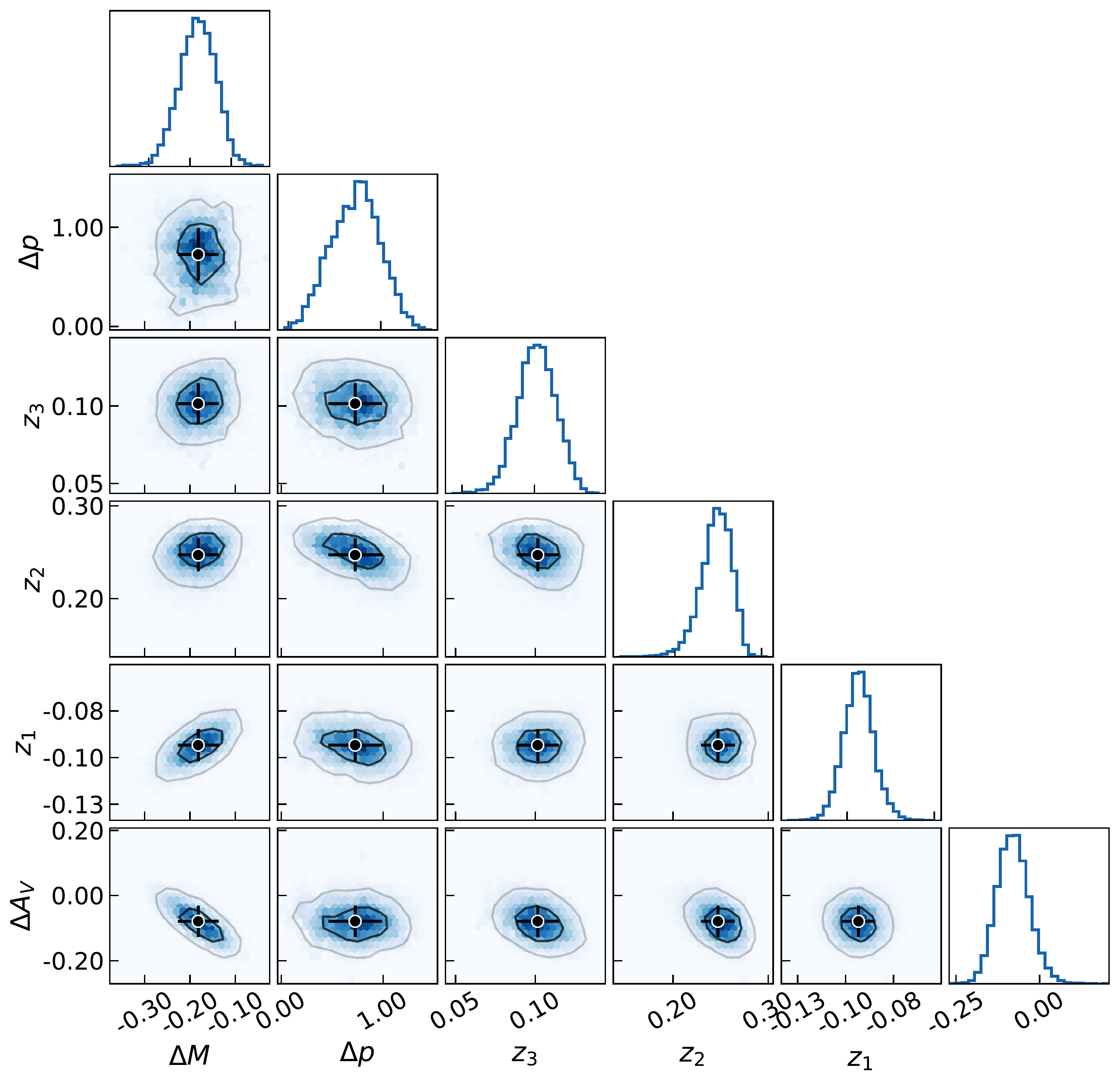}

\caption{Top: PAE reconstruction (red) and best fit SALT2 model (blue) of two supernovae from the test set (black). For visualization purposes the spectra in the top panels have been shifted vertically by a constant factor of the observation time. Bottom: Corresponding best fit PAE model parameters and their errors determined from Hamiltonian Monte Carlo.}
\label{fig:testing_spectra}
\end{figure*}

First, in Sections~\ref{sec:vary_latent}~and~\ref{sec:reconstructions} we look at the spectral features captured by the PAE parameters and the reconstruction accuracy of our PAE model in comparison to the SALT2 model. In Section~\ref{sec:generative} we discuss the straightforward generation of simulated supernovae observations consistent with the data distribution. Followed by a search for any outlying supernovae in Section~\ref{sec:outliers}, and the determination of cosmological distance accuracy in Section~\ref{sec:PAE_results}.

\subsection{Latent parameters to spectral variations}
\label{sec:vary_latent}

The supernovae spectral time series we are attempting to reconstruct have an overall consistent shape at a given rest frame time, with small variations from object-to-object. Therefore the observation time fed to the decoder will determine the time evolution of the supernovae, and variations in each latent parameter describe the object-to-object spectral variability encoded by a combination of those dimensions. 

In Figure~\ref{fig:vary_latent} we demonstrate how separately varying each latent parameter from their mean values in a three non-linear latent dimensional model affects the reconstructed spectra. We find that the latent parameters have each encoded unique spectral information. The first and second dimensions by design were restricted to learn physical components of the model, where the former dimension encodes the extrinsic amplitude $\Delta M$, and the latter is the time-independent color-law relative extinction coefficient $\Delta A_V$. The remaining dimensions are free to learn any spectral variations that exist in the SNe~Ia population used for training. For the specific model shown here we find that the $z_1$ dimension seemingly resembles a combination of an amplitude multiplication correlated with certain absorption/emission features and a brighter-slower effect. The $z_2$ and $z_3$ intrinsic latent dimensions focus more on details of the absorption/emission features and spectral tilt. We note that unlike a PCA decomposition where components are ranked by the variance they explain, our autoencoder has no such constraint, and the intrinsic latent parameters are free to learn any modes of spectral diversity. The fact that the first intrinsic latent parameter happened to result in the most apparent modifications to the reconstructed spectra is a coincidence.

The key difference between our non-linear PAE model and a linear PCA analysis is that the latent-dimensions of the PAE are both non-independent and non-symmetric around the mean. We can see clearly from Figure~\ref{fig:vary_latent} that the effects of a latent value smaller than the mean (blue) are not simply the inverse of a latent value larger than the mean (red), but describes independent information. This allows for more information to be encoded within a single dimension in comparison to PCA, where each dimension is simply a multiplier in front of a tempo-spectral component (i.e. $x_1 M_1(t, \lambda)$ in the SALT2 model). Additionally, as the latent parameters are passed through a number of non-linear layers of the decoder, their effects on the reconstructed spectra are not limited to the spectral variations of the independent $z_1$ and $z_2$ dimensions, but can interact in highly non-linear ways to produce more complex spectral features than those shown here. We show a model with three non-linear parameters for visualization purposes, but the method can be increased to any dimensionality. 

\subsection{Accuracy of PAE data reconstructions}
\label{sec:reconstructions}

\begin{figure*}
\centering
\includegraphics[width=1.0\textwidth, trim=0 90 0 0, clip]{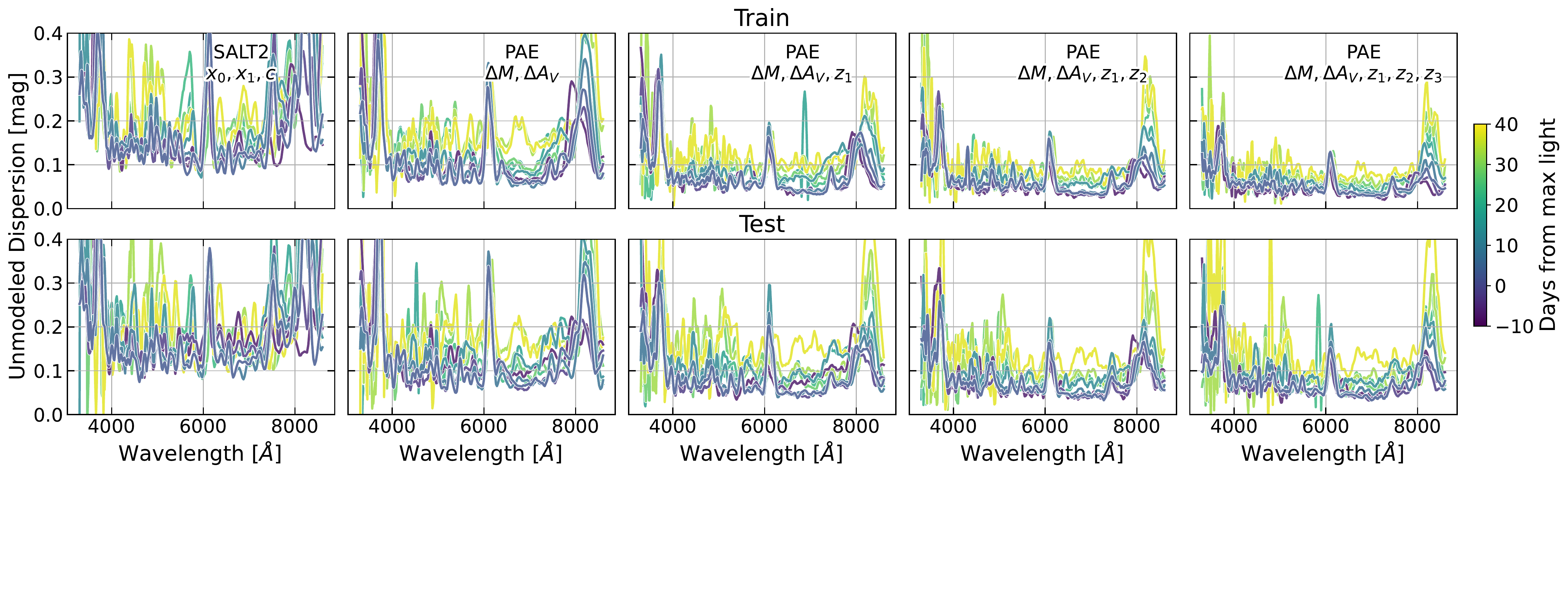}
\caption{Unmodeled dispersion -- the additional dispersion beyond the observational uncertainty required to explain the variance of the reconstructions and the data (Equation~\ref{eq:intrinsic_magnitude}) -- of SALT2 and our PAE model with increasing latent space dimensionality. The dispersion is measured in five day intervals for the training data (top) and on the unseen test data (bottom). Beyond 3 non-linear dimensions ($z_1, z_2, z_3$), plus extinction ($\Delta A_V$) and a free amplitude scaling parameter ($\Delta M$), we found no improvement on the test sample, and thus do not display the additional panels here.}
\label{fig:mag_disp}
\end{figure*}

\begin{figure*}
\centering
\includegraphics[width=0.6\textwidth, trim=0 0 90 0, clip]{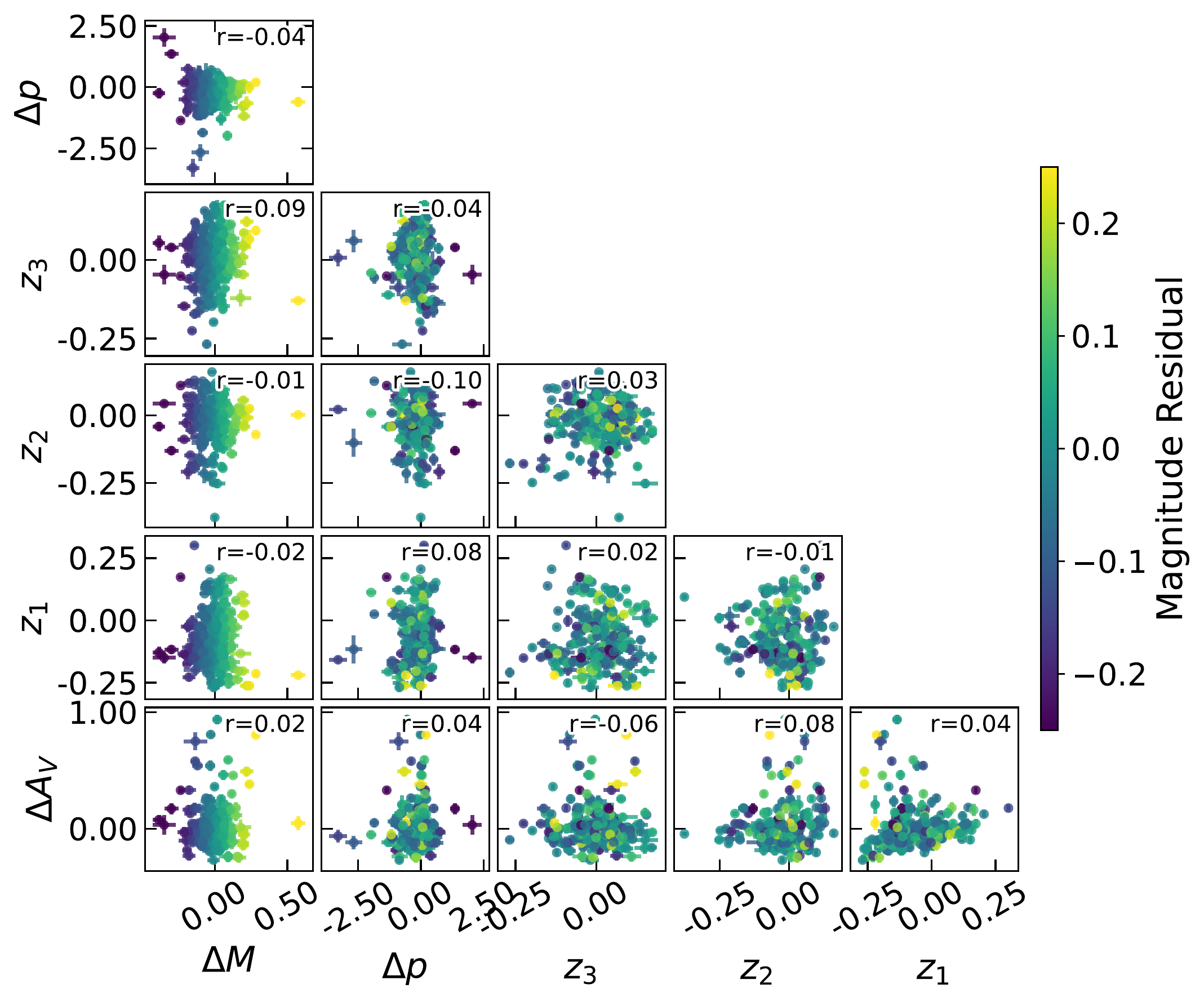}
\caption{Best fit PAE parameters for all supernovae with a redshift greater than 0.02. The Pearson correlation coefficient $r$ is shown in the top right of each panel.}
\label{fig:parameters}
\end{figure*}

While we demonstrated that varying latent parameters of the PAE captures a number of complex spectral and temporal features, the key for using the model is its accuracy of modelling the SN~Ia observations in the training and test sets. In Figure~\ref{fig:testing_spectra} we compare the SALT2 and PAE reconstructions of the data for two supernovae from the test set. We chose to display a supernova with many observed spectra and a low level of observational noise (left), and a supernova with only a few observations, with none before peak brightness, and an increased observational noise level (right). These two examples demonstrate the diversity of objects in the dataset, and are a representative display of the performance of both the SALT2 and PAE models. 

We find that the PAE reconstructions are highly accurate over the entire observation range from $-10$ to $+40$ days, even for samples that have highly non-uniform time sampling. In comparison to the SALT2 best fit spectra, we see a better fit overall, both on the amplitude offset, and on the matching of absorption and emission features on the spectra, particularly the CaII H\&K, Si II, and OI features at $\sim$3950 \AA, $\sim$6150~\AA, and $\sim$7800~\AA, respectively. For SNe~Ia with abnormally large luminosity at early times (e.g., \citet{nordin2018}) we find that the PAE reconstruction still matches the observations to high accuracy, while the SALT2 model fails to capture the spectral diversity of these types. 

The accuracy of the reconstructions of the PAE model depends on the number of latent dimensions used. Too few dimensions does not allow for the full spectral variability of the supernovae time series to be expressed, while too many dimensions can allow the model to improve the fit on the training data, to no improvement of the test data. By training multiple autoencoders, each with a different number of intrinsic non-linear latent parameters, we studied the optimal latent dimensionality for reconstruction quality. Using the same AE architecture and multi-stage training procedure described in Section~\ref{sec:architecture} we varied the dimensionality of the latent space from two to eight. When referring to the dimensionality of the latent space we count only the model parameters that capture intrinsic and extrinsic effects, and do not include the time shift relative to the SALT2 fits, $\Delta p$.

To quantify the quality of the model reconstructions we report the level of unmodeled dispersion -- the additional dispersion beyond the observational uncertainty required to explain the variance of the reconstructions and the data. This is determined by modelling the observed flux $f_{\mathrm{obs}}$ as 
\begin{equation}
    \label{eq:intrinsic_magnitude}
    f_{\mathrm{obs}}=\NN(f_{\mathrm{model}}, \sigma^2_{\mathrm{obs}} + \sigma^2_i),
\end{equation}
and fitting for the maximum likelihood of the unmodeled dispersion $\sigma_i$. We report this value in magnitudes for each wavelength, binned in 5 day intervals, and show the results as a function of latent space dimensionality in Figure~\ref{fig:mag_disp}. 

We find that our PAE outperforms the standard SALT2 model at all wavelengths and observation times, and that increasing the latent dimensionality continues to decrease the unmodeled dispersion across the time and wavelength range up until three non-linear latent parameters ($z_1, z_2, z_3$), after which it flattens to show no additional improvement on the test set. The dispersion near the CaII H\&K and Si II lines ($\sim$3950~\AA, $\sim$6150~\AA), and near the Ca NIR triplet at $\sim$8100~\AA, shows the most significant improvement when increasing the non-linear latent dimensionality. This clearly demonstrates that additional components describing the intrinsic variations of the SN~Ia population can learn increasingly complex spectral and temporal features. The dispersion between $\sim$6500 to $\sim$7750~\AA\  remains at $\sim0.05$ magnitudes, as this region has little to no spectral features that vary between supernovae. We find that the unmodeled magnitude dispersion near peak brightness is on average the lowest, and increases at later times. Between 30 and 40 days after peak brightness we find that the dispersion is larger than near the peak. The dispersion is higher even in spectral regions not associated with strong spectral features, suggesting that the uncertainty is somewhat under-estimated for these very faint spectra.
Integrating the test set over a B-band bandpass we find that the SALT unmodeled dispersion near peak brightness is 0.128\,mag, compared to a PAE value of 0.056\,mag -- a factor of 2.28 larger.

\begin{figure*}
\centering

\includegraphics[width=1.0\textwidth, trim=0 0 200 0,clip]{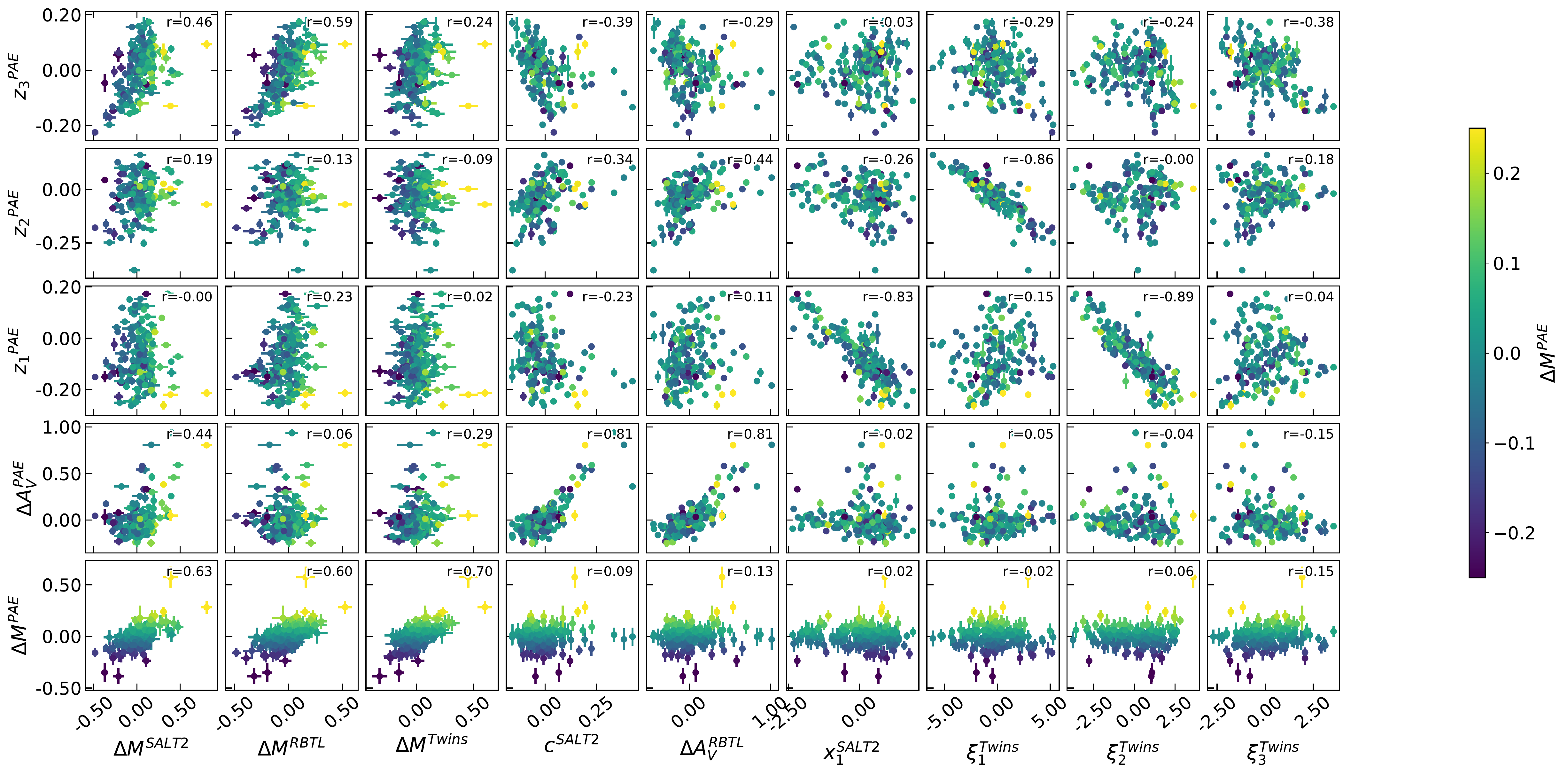}

\caption{PAE parameters compared to SALT2 and the Twins Embedding/RBTL for overlapping supernovae with a redshift greater than 0.02. The Pearson correlation coefficient $r$ is shown in the top right of each panel.}
\label{fig:parameters_vs_others}
\end{figure*}

We select the three non-linear latent dimension model as optimal for modelling the data, and as we will show below it also returns the lowest magnitude residuals. We examine the best fit model parameters in Figure~\ref{fig:parameters}. We note that the parameter values shown are those found by finding the minimum of the log posterior through Hamiltonian Monte Carlo, and not simply the encoded values of each supernova. This ensures that the full parameter space has been explored, and thus the variations of any parameter are not artificially small due to any limitations of the encoder. From visual inspection we find that the magnitude residual contains no noticeable correlations with the other model parameters, confirming that our multi-stage training setup and correlation penalty has ensured that the intrinsic model parameters have learned clear correlations between intrinsic luminosity and spectral and/or temporal features of SNe~Ia. Given that the dimensionality of the non-$\Delta M$ parameters is large, it is possible that small correlations between these parameters and the magnitude residual remain. If so, these correlations could be uncovered with an additional non-linear model, which could be then be used to explain and reduce the extrinsic magnitude such as in the SALT2 or the Twins Embedding analysis. Initial investigations with a fully connected neural network trained on the latent parameters of the training set did not reduce the extrinsic magnitude dispersion when applied to the test set. We also find that the average time shift relative to the SALT2 fit, $\Delta p$, is within approximately half a day and is consistent with the uncertainty of the SALT2 fits, and that the standard deviation of $\Delta A_v$ is 0.132.  A small number of supernovae have time shifts of a few days relative to the SALT2 best fits. These are mostly supernovae with no observations near or before peak brightness. 

Figure~\ref{fig:parameters_vs_others} compares the best fit PAE parameters to the SALT2 and the Twins Embedding models described in Section~\ref{sec:data}. As expected, we find a high degree of correlation between the magnitude residuals ($\Delta M^{SALT2}, \Delta M^{Twins}, \Delta M^{PAE}$) and the color ($c^{SALT2}, \Delta A_V^{RBTL}, \Delta A_V^{PAE}$) between the three models, although there is a non-zero scatter. We find that the magnitude residuals cover a similar range of values, while the relative extinction inferred by the Twins Embedding covers a larger range of values than that of our PAE. This larger range is likely due to the multiple steps required to perform the Twins Embedding. Rather than train all parameters simultaneously as for the PAE, the Twins Embedding magnitude and extinction are fit first in the two-parameter RBTL step, and thus $\Delta A_V^{RBTL}$ is forced to simultaneously explain both the extinction and any intrinsic color-like effects. Alternatively, the PAE simultaneously learns all intrinsic and extrinsic parameters, and the intrinsic latent parameters ($z_1, z_2, z_3$) can learn any color-like features that happen to be correlated with spectral or temporal features. If we instead allow the PAE to learn only a magnitude and extinction, we find that the best-fit extinction values are nearly equivalent to those reported by the Twins Embedding.  

We find a clear correlation between our $z_1$ parameter and the $x_1$ parameter of SALT2, and various correlations between our latent space and the latent Twins Embedding parameters ($z_1 \rightarrow \xi_2, z_2 \rightarrow \xi_1, z_3 \rightarrow \xi_3$).      

\subsection{Simulating new SNe~Ia}
\label{sec:generative}

The generation of new supernova samples consistent with the data is straightforward with a probabilistic autoencoder. We simply sample a random latent vector $\uu$ from a unit Gaussian, pass this through the normalizing flow to get the sample in autoencoder latent space $\zz$, append the desired observation times and magnitude offset $\Delta M$, and pass this through the decoder to yield a new spectral time series. By the probabilistic nature of the normalizing flow Gaussian latent space $\uu$, the distribution of generated samples corresponds to the density of similar samples in the training dataset -- significant outliers will be rare, while ``average'' spectra will have a higher probability of being generated.

\subsection{Detecting Outlying Supernovae}
\label{sec:outliers}

\begin{figure*}
\centering
\includegraphics[width=0.55\textwidth]{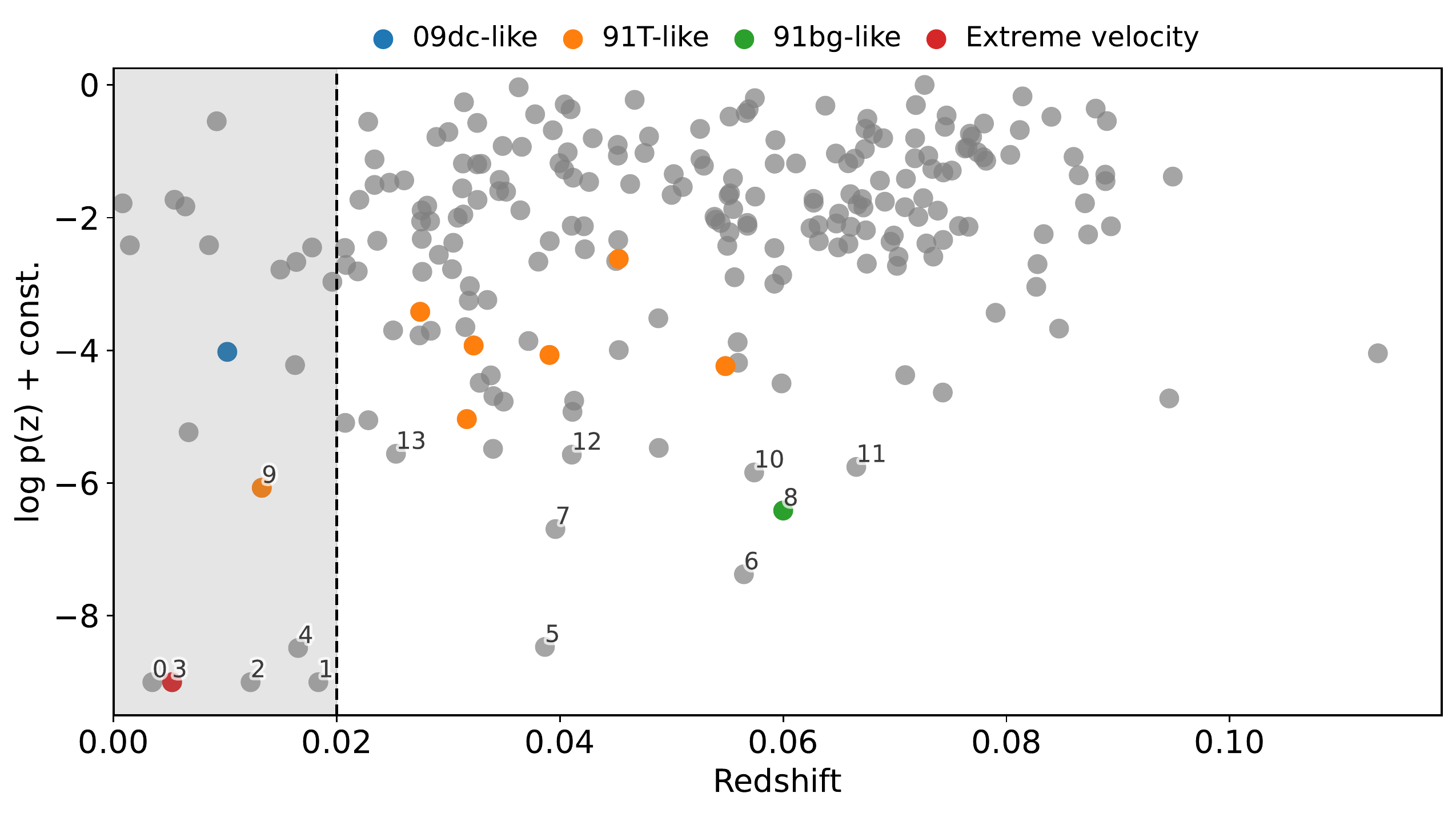}
\includegraphics[width=0.44\textwidth]{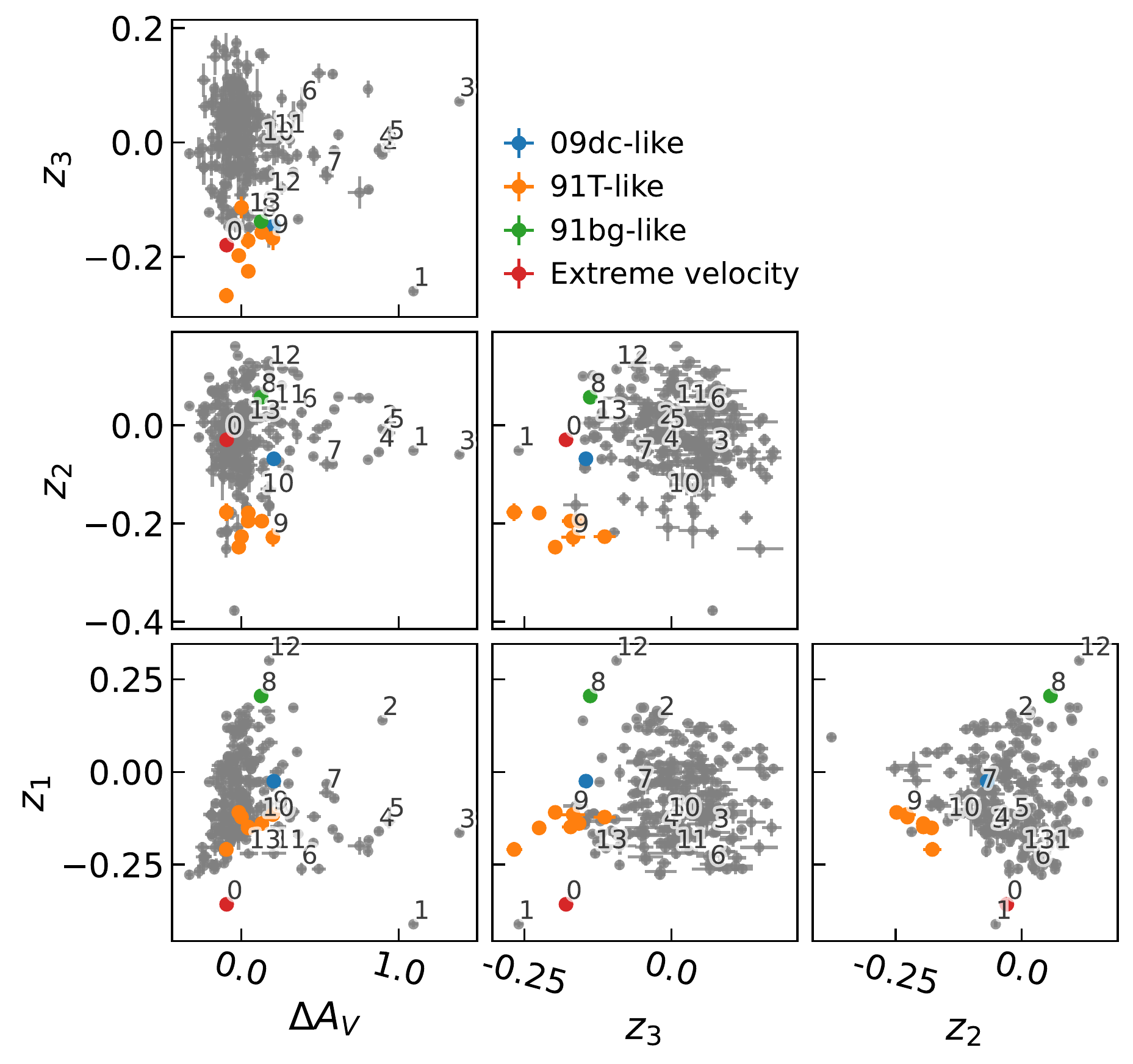}

\caption{Best fit PAE latent parameters with peculiar SNe~Ia displayed as colored markers. The left panel shows the density of each supernova as a function of the redshift, while the right panel displays where the supernova resides in each latent dimension. We annotate the 14 lowest density examples in order to enable comparisons between the left and right panels.}
\label{fig:outliers}
\end{figure*}

The PAE framework allows for an effortless determination of data density in both $\zz$ and $\uu$ space, which are simply related through the Jacobian determinant of the normalizing flow. Samples residing in low density regions of the latent space are less similar to other supernovae, while those in high density regions are more similar to others. Therefore, by selecting supernovae by the density of their latent representations, $\mathrm{p(\zz)}$, we can pull out outliers or common samples for further inspection. Low density does not mean that these supernovae are more poorly fit by the PAE -- we do not find that either the reconstruction error or magnitude residual depends on density -- only that the latent parameters and hence corresponding spectral characteristics are more unique. 

Figure~\ref{fig:outliers} shows the density as a function of redshift for all supernovae (left), calculated from their best-fit latent parameters. We find no distributional shifts between supernovae in the training and test sets, so we include all 228 supernovae here. It is clear that there are a few supernovae residing in low density regions of the latent space - i.e. with the smallest values of $\mathrm{log\ p(\zz)}$. $\Delta A_V$ is included, and because strongly reddened SNe~Ia can easily become isolated in that dimension, low values for $\mathrm{log\ p(\zz)}$ can result. This explains all the cases with $\mathrm{log\ p(\zz)} < 10^{-8}$. A number of these are at low redshifts, but as we have excluded the amplitude parameter $\Delta M$ from the density calculation as described in Section~\ref{sec:posterior_methods}, the density estimation should immune to the effects of peculiar velocities on amplitudes.

For a small number of supernovae in our dataset we have external labels specifying peculiar sub-types, including 91T-like, 91bg-like, and 09dc-like, which we highlight with different colored markers. We find that the best fit PAE model parameters for these supernovae are in regions of lower than average density, which is expected given that they belong to rare sub-populations. As the number of supernova of a certain sub-type grows their region of the latent space becomes well-populated, and thus a low-likelihood is more effective at finding individual rare SNe, such as the 91bg-like example, rather than whole sub-populations such as the 91T-like. For sub-populations it is better to examine clusters in various regions of the latent space.

A closer inspection of individual dimensions of the latent space (right panel) shows that the low density supernovae are not necessarily peculiar among all dimensions, rather their peculiar features are often isolated in a specific dimension, such as large values of $\Delta A_V$, as noted above, or low values of $z_2$ or $z_3$. We annotate the 14 lowest density examples in order to enable comparisons between the left and right panels. We find that five of the six lowest density examples are supernovae with high extinction, and are mostly at low redshifts. The large extinction results in a relatively small transmitted flux, so perhaps similar SNe~Ia at high redshift are simply below the flux detection threshold. We also find a clear cluster of 91T-like supernovae with low values of both $z_2$ and $z_3$, demonstrating that the PAE is a valuable tool for population studies of supernovae. Beyond the five low redshift supernovae with high extinction we find no clear population shifts as a function of redshift. 


\subsection{Cosmological Distance Measurements}
\label{sec:PAE_results}

\begin{figure*}
\centering
\includegraphics[width=1.0\textwidth]{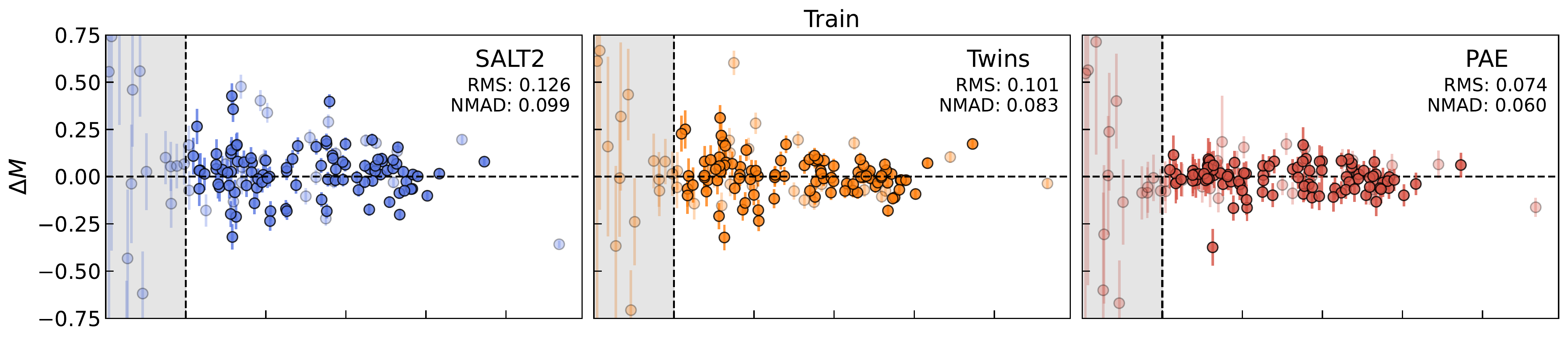}
\includegraphics[width=1.0\textwidth]{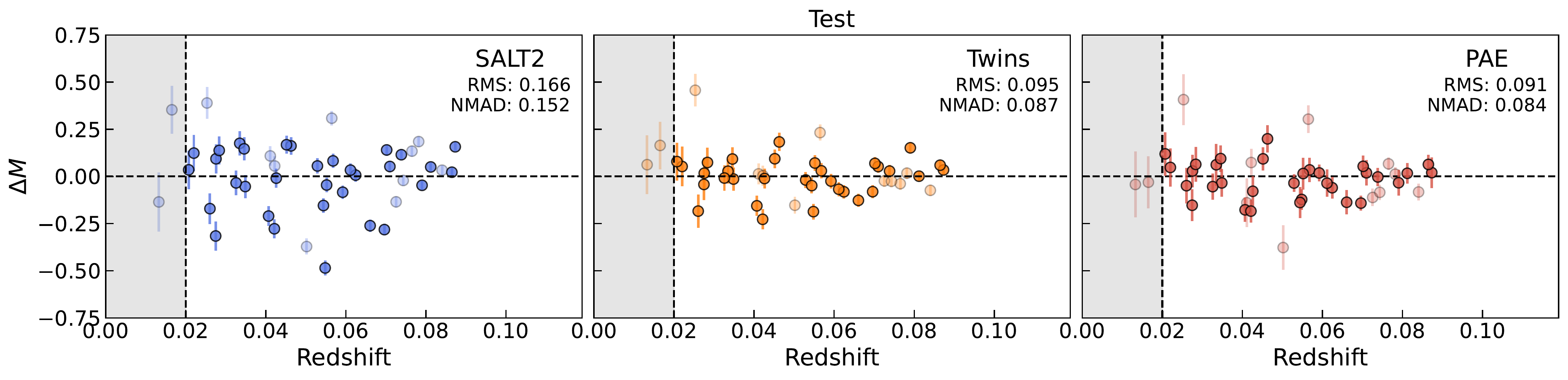}

\caption{Magnitude residuals of supernovae from the training (top) and test (bottom) sets, including the component from peculiar velocities. Both the SALT2 and Twins Embedding results are obtained from a linear (SALT2) or non-linear (Twins) magnitude standardization procedure, while our probabilistic autoencoder has been trained to explicitly separate the extrinsic magnitude from the intrinsic supernovae luminosity, and thus requires no standardization. Solid data points are those used in the final Twins Embedding cosmological distance analysis, while data points with some transparency are the remaining overlap between our data and the full Twins Embedding data set.}
\label{fig:mag_redisuals_2}
\end{figure*}

\begin{table*}
    \centering
    \begin{tabular}{l r c c c}
    \toprule
    Data Sample & Statistic & SALT2  & Twins Embedding & PAE \\
    \hline
    & NMAD      &  $0.099 \pm 0.013$ &   $0.083 \pm 0.014$ &  $0.060 \pm 0.010$ \\
    Train & RMS       &  $0.126 \pm 0.012$ &   $0.101 \pm 0.009$ &  $0.074 \pm 0.010$ \\
    & RMS w peculiar velocity removed &  $0.115 \pm 0.012$ &   $0.087 \pm 0.009$ &  $0.052 \pm 0.010$ \\
    
    \hline
    
    & NMAD      &  $0.152 \pm 0.039$ &  $0.087 \pm 0.019$ &  $0.084 \pm 0.023$ \\
   Test & RMS       &  $0.166 \pm 0.022$ &  $0.095 \pm 0.012$  &  $0.091 \pm 0.010$ \\
    & Peculiar velocity removed &  $0.157 \pm 0.022$ &  $0.079 \pm 0.012$ &   $0.074 \pm 0.010$ \\

    \end{tabular}

    \caption{Standardization performance for all methods presented in this work, showing the NMAD, RMS, and an estimate of the RMS with the peculiar velocity removed (see text for details). The significant difference between the train and test performance of the SALT2 model stems from the random assortment of supernovae into each set, which combined with the small number of total samples happened to place a higher fraction of large SALT2 magnitude residual SN into the test set. This is reflected in the increase in uncertainty reported on the RMS and NMAD, determined by bootstrap resampling \citep{bootstrap}.}
    \label{tab:rms}

\end{table*}
The application with perhaps the most scientific utility is the determination of the magnitude residual for each supernova, which is the key factor for determining the distance accuracy. As explained in Section~\ref{sec:architecture} the three-stage training and correlation penalty term encouraged our PAE model to separate the extrinsic magnitude component, which is uncorrelated with features of the spectral time series, from any amplitude-like modification that is correlated with intrinsic spectral features or temporal evolution. Thus the extrinsic magnitude residual $\Delta M$ is directly fit for during the posterior analysis phase of our analysis, and we do not require any additional steps to uncover correlations between model parameters and magnitude residuals in order to perform magnitude standardization. This differs from the methodology employed in the SALT2 and Twins Embedding models that we compare to, which employ an additional linear (SALT2) or non-linear (Twins) model to predict the magnitude given the model parameters, and remove this predicted value to obtain a final magnitude residual. Although the PAE model parameters may end up having some small remaining correlations with $\Delta M$ which can be exploited to explain the magnitude residuals, we do not perform any magnitude standardization for the results shown here. Whether magnitude standardization benefits from learning correlations during model training, rather than determining them after model training is completed, is not obvious {\it a priori}, but is nevertheless an interesting topic for future investigation. 

As outlined in Section~\ref{sec:posterior_methods} we marginalize over ($\Delta M$, $\uu$, $\Delta p$) for each SN~Ia to obtain the final best fit model parameters and their uncertainty through Hamiltonian Monte Carlo starting from the best fit MAP parameter, using the mean and percentiles of the posterior samples as the mean and error on the best fit parameters. To enhance the predictive performance and error estimation on the physical supernova parameters inferred by a single PAE model we use the weighted mean and variance calculated from the results of 10 separate models. Each model is trained in an identical fashion on the training set with the procedure described above, but using a different random seed to initialize network weights. On the $\Delta M$ parameter for all models we include a peculiar velocity error component, assuming a velocity of $v_{\mathrm{pec}} = 300$~km/s\footnote{This is the prediction from linear perturbation theory \cite[see, e.g., ][]{hui2006}}. 

We first studied the magnitude residuals as a function of the latent dimensionality of the model, finding a clear decrease in the magnitude residual for both the training and test sets as we increase the latent dimensionality from 0 to 3 non-linear intrinsic parameters. Similar to the intrinsic dispersion results of Figure~\ref{fig:mag_disp} we find no statistically significant improvement when increasing beyond a model with three non-linear parameters (i.e. ($z_1, z_2, z_3$) plus extinction and an amplitude scaling). This is in agreement with \citet{twins1, twins2}, who find quickly diminishing returns when expanding beyond three non-linear intrinsic model parameters. From this investigation we determined that this model is optimal for explaining the diversity of SNe~Ia given our dataset, and again restrict to this model for the following results.

In Figure~\ref{fig:mag_redisuals_2} we compare the PAE magnitude residuals to those derived by SALT2 and Twins Embedding analyses for SNe~Ia in common -- 137 SNe in the training set and 44 in the test set. Of this overlapping fraction 96 of the training and 32 of the test were part of the final Twins cosmological distance analysis. We display both the (unweighted) root mean squared (denoted $\sigma$) and the Normalized Median Absolute Deviation (NMAD) of the magnitude residuals. While both statistics are similar, the NMAD is less susceptible to large outliers. We show the RMS and NMAD for SALT2, Twins, and PAE over the subset of supernovae that overlap with the ones used in the final Twins Embedding cosmological distance analysis. 

It is important to note that each analysis had somewhat different subsets of SNe~Ia available for development and training, especially given the small number of data samples available which theoretically could cause a small number of outlying supernovae that exist in one dataset but not another to considerably alter model training. Thus although this analysis facilitates a comparison of the magnitude standardization capabilities of the three models, the magnitude residuals we report inevitably reflect a combination of the strength of the method at explaining the diversity of SNe~Ia coupled with signatures of the specific data used for model training -- it is impossible to disentangle model implementations from subtle effects introduced by the different training data. However, by focusing only on the supernovae that do overlap between the different analyses, we attempt to mitigate any such differences.  


For individual SNe~Ia we see consistent results across all models, specifically the low redshift objects with large magnitude residuals resulting from significant peculiar velocities, which helps to validate the fitting procedure employed in our analysis. Our model was trained using a minimum redshift cut of 0.02, and thus was was not trained using any objects with significant redfhift contribution due to peculiar velocity, but by excluding the $\Delta M$ parameter from the normalizing flow we have included no prior on the extrinsic amplitude and it is allowed to freely vary to any value which best fits the data.

We find that our PAE obtains significantly smaller magnitude residuals than the SALT2 model, and shows a magnitude residual similar to that of the Twins Embedding analysis. None of the SNe in our dataset were used for training the SALT2 model, such that the test set illustrates an unseen sample and thus a true test sample performance for both SALT2 and our PAE. The samples we display for Twins will include both samples used for training and those used for testing. For the PAE we find a smaller magnitude residual for the training set in comparison to the test set, which is commonly found in the generalization of deep neural networks to unseen samples, but the difference is not statistically significant. This does not equate to model overfitting, in which continuing to improve the fit on the training data comes at a cost of decreasing performance on unseen test data. As discussed in Section~\ref{sec:training} we find no evidence for this, and the test error continues to decrease until flattening. Had we a larger training sample we would have separated out a validation set, and stopped training when the error on this set stopped decreasing, but we did not want to further reduce the amount of available training samples by separating out a validation set in addition to the test set. This choice of not performing early stopping during training was made to ensure that the model remained blind to the test set.

We tabulate the final magnitude residuals for all methods in Table~\ref{tab:rms}. While we do not have the same supernovae in our dataset that were used in \citet{twins2}, we find that the statistics we report for their results are similar to those quoted in their paper of $\mathrm{NMAD} = 0.83 \pm 0.010$ and $\sigma = 0.101 \pm 0.070$ over the full sample of 134 SNe that passed their data cuts. To approximate removing the component of the magnitude residual stemming from peculiar velocities we assume a 300~km/s velocity for each SN, which contributes an added dispersion of 0.053 mag. We subtract this from the RMS in quadrature. The uncertainty reported on the RMS and NMAD are determined by bootstrap resampling \citep{bootstrap} of the magnitude residuals.

The agreement between the Twins Embedding analysis and our PAE is interesting given that the data products used to perform the analysis are quite different. The Twins Embedding models the diversity of SNe~Ia only at maximum light, interpolating a max light spectra using observations from within ($-5$ days, $+5$ days) from peak brightness. Therefore any information from observations outside of this window is not included in the Twins Embedding analysis, while the PAE is allowed to learn the full temporal evolution in addition to the diversity at maximum light. Nevertheless, the magnitude residuals between the two methods are similar, seemingly pointing towards observations near peak brightness being the most important for standardization, similar to as noted by \citep{fakhouri2015, sugar}. To investigate the PAEs reliance on observations near peak we perform an equivalent posterior analysis while masking any spectra within a ($-5$ day, $+5$ day) range from maximum brightness. We find that when only using spectra outside of this range the NMAD and RMS increase to $0.080 \pm 0.014$ and $0.112 \pm 0.011$ on the training set, and $0.092 \pm 0.032$ and $0.135 \pm 0.017$ on the test set. This magnitude residual when only using observations away from peak brightness, while still an increase over using the full timeseries, demonstrates that the PAE does not require any observations near peak brightness to still obtain relatively small magnitude residuals, and implies that the intrinsic brightness of the supernovae can be determined from spectra at any date from $-10$ to $+40$ days relative to peak brightness.

\section{Discussion \& Conclusions}
\label{sec:discussion}

The goal of this paper is to develop a single framework for SN~Ia data analysis, developing a data driven model that can be used for all of the downstream tasks, including posterior analysis of all of its parameters including the distance modulus, anomaly detection, and realistic SN~Ia spectro-temporal simulations. Our approach is a physically-parameterized probabilistic autoencoder (PAE) to model type Ia supernovae spectral evolution. We showed that the model, trained directly on the data without any data cuts, separately learns both intrinsic variation and extrinsic variation (dust and distance modulus) of supernovae variability, and can model the data to very high accuracy. We introduced a multi-stage training procedure, which with the addition of a correlation penalty term between the model parameters, disentangles extrinsic magnitude changes due to peculiar velocities from the portion of intrinsic luminosity of the supernovae that correlated with optical spectral and/or temporal features.

The disentanglement of intrinsic and extrinsic effects during training is novel
to this work. Usually the discovery of correlations between model parameters and magnitude dispersion composes an additional ``magnitude standardization'' step after the model is trained, which requires another linear or non-linear model, and thus introduces another set of errors that need to be propagated through to the final constraints. In contrast, in our approach all of the training is done once. We demonstrate that the intrinsic scatter, even when inflated by the peculiar velocity dispersion, can be as low as 0.1 magnitude, which bodes well for peculiar velocity measurements with local supernovae.

Both our analysis and that of \cite{twins2} show that SNe~Ia inhibit a 3-dimensional parameter space (see \citet{rubin2020} for a more complete discussion of this concept). Physical modeling of SN~Ia explosions involve many more parameters \cite[e.g.,][]{hillebrandt2000}, so our results imply the presence of strong correlations among these parameters. The $z_{1,2,3}$ space presented here offers an efficient means of comparing SN~Ia model results with real SNe~Ia. Significant challenges remain in producing high-fidelity spectral models of SN~Ia explosions \citep[cf.][]{roepke2012}, as even small changes in the modeling of radiative transfer produce strong effects on model spectra. More mature physical models can eventually be efficiently compared in our $z_{1,2,3}$ space.

We release all codes, and the trained models, at \href{https://github.com/georgestein/suPAErnova}{github.com/georgestein/suPAErnova}.

\section{Acknowledgements}
\label{sec:acknowledgements}
MR has received funding from the European Research Council (ERC) under the European Union's Horizon 2020 research and innovation programme (grant agreement n°759194 - USNAC)
This work was supported in part by the Director, Office of Science, Office of High Energy Physics of the U.S. Department of Energy under Contract No.~DE-AC02-05CH11231. Support in France was provided by CNRS/IN2P3, CNRS/INSU, and PNC and French state funds managed by the National Research Agency within the Investissements d'Avenir program under grant reference numbers ANR-10-LABX-0066, ANR-11-IDEX-0004-02 and ANR-11-IDEX-0007. Additional support comes from the European Research Council (ERC) under the European Union's Horizon 2020 research and innovation program (grant agreement No 759194-USNAC). Support in Germany was provided by DFG through TRR33 "The Dark Universe" and by DLR through grants FKZ 50OR1503 and FKZ 50OR1602. In China support was provided from Tsinghua University 985 grant and NSFC grant No~11173017. Some results were obtained using resources and support from the National Energy Research Scientific Computing Center, supported by the Director, Office of Science, Office of Advanced Scientific Computing Research of the U.S. Department of Energy under Contract No. DE-AC02-05CH11231. We also thank the Gordon \& Betty Moore Foundation for their support.

\bibliographystyle{aasjournal}
\bibliography{bibliography}

\newpage

\pagebreak

\clearpage

\label{lastpage}
\end{document}